\documentclass[preprint,12pt]{elsarticle} 
\usepackage{textcomp}
\usepackage{amssymb} 
\usepackage{amsmath} 
\usepackage{xcolor}
\usepackage{gensymb}
\usepackage[nolist]{acronym} 
\usepackage[mathlines]{lineno} 
\usepackage{booktabs}
\usepackage{physics}%
\usepackage[hyphens]{url}
\usepackage{ulem}
\pdfoutput=1
\usepackage{graphicx}
\usepackage{soul}

\begin{document}

\begin{frontmatter} 

\title{Radio wavefront of very inclined extensive air-showers: a simulation study for extended and sparse radio arrays}
\author[psu,psuigc,iap]{Valentin~Decoene}
\ead{decoene@iap.fr} 
\author[lpnhe,iap]{Olivier~Martineau-Huynh}
\author[iflp,iap]{Matias~Tueros}
 
\address[psu]{Department of Physics, The Pennsylvania State University, USA, University Park 16802 PA}
\address[psuigc]{Center for Multimessenger Astrophysics, Institute for Gravitation and the Cosmos, The Pennsylvania State University, USA, University Park 16802 PA}
\address[iap]{Sorbonne Universit\'e, CNRS-UMR 7095, Institut d'Astrophysique de Paris, 98 bis boulevard Arago, F-75014 Paris, France}
\address[lpnhe]{Sorbonne Universit\'e, CNRS/IN2P3, Laboratoire de Physique Nucl\'eaire et de Hautes Energies, LPNHE, 4 place Jussieu, F-75005 Paris, France} 
\address[iflp]{Instituto de F\'isica La Plata - CONICET/UNLP, Diagonal 113 entre 63 y 64 La Plata (1900) - Buenos Aires, Argentina}

\begin{abstract} 
Radio-detection is becoming an established technique for the detection of air showers induced by cosmic particles. This is in particular true at the highest energies, where very large detection areas are required. A proper description of the shape of the radio wavefront emitted by air showers may allow to reconstruct the properties of its parent particle. In this article, we show that for showers with zenith angles larger than 60$\degree$ ---those targeted by giant radio arrays detecting extensive air showers induced by cosmic particles---, a point-source-like description of the radio wavefront allows to constraint the lateral position of the shower axis within a few meters.

Following, we show that the reconstructed longitudinal position of this point source is correlated with the nature of the cosmic rays initiating the shower. Further systematic studies are pending to determine the robustness of this parameter and its validity as a proxy for cosmic ray composition studies.
\end{abstract}

\end{frontmatter}

\section{Introduction}
A proper description of the shape of the electromagnetic wavefront emitted by an extensive air shower can be a key tool to reconstruct the properties of the cosmic particle at the origin of the air shower. 

Substantial modeling efforts and experimental work have already been carried out on this issue~\cite{SCHRODER20171}. The LOPES and LOFAR collaborations have in particular studied in detail the radio wavefronts from vertical air showers observed with their detectors, and concluded that a hyperbolic shape describes the data best~\cite{Apel_2014, Corstanje_2015}.

In this paper, we focus on the specific case of inclined showers, whose wavefront may differ from those observed by LOPES or LOFAR because they develop farther from ground and induce much larger footprints. This work has a particular interest in the framework of the GRAND project~\cite{Alvarez-Muniz:2018bhp}, a proposal to build a giant network of radio arrays aiming primarily at detecting UHE cosmic neutrinos through the nearly horizontal showers they induce in the atmosphere. This is also relevant for the BEACON project, which aims at the same goals but using phased antenna stations in the $30-80$\,MHz range and deployed atop of high altitude mountains~\cite{Wissel_2020}.

We study in section \ref{sec:study} the shape of the radio wavefront from air shower simulations carried out over a three-dimensional antenna array. The defined treatment is then applied in section \ref{sec:recons} to inclined showers illuminating a simulated layout deployed on the ground. The potential of this method in terms of cosmic-ray composition determination is finally showcased.

\section{Study of the wavefront curvatures}
\label{sec:study}
\subsection{Wavefront description}
\label{sec:pheno}
The instant $t_i$ (i.e., antenna trigger time)  when the wave passes at a position $\vec{x_i}$ (i.e., antenna position) can be described in the general case as the sum of two terms: a pure propagation term $\mathcal{P}$ and an intrinsic curvature term $\mathcal{C}$, simply given by
\begin{align} \label{eq:wvf:general}
    c\,t_i\qty(\vec{x}_i) = \mathcal{P} + \mathcal{C} \ ,
\end{align}
where $c$ is the velocity of light. In a phenomenological approach, the propagation term $\mathcal{P}$ can be assimilated to the linear translation of a plane wave along the propagation direction at light speed from the emission region down to the antenna position
\begin{align} \label{eq:wvf:plane}
    \mathcal{P} = n_i \vec{k}\cdot\vec{x_i},
\end{align}
where $\vec{k}$ is the unit direction vector of the shower, and $n_i$ the mean value of the refractive index along the line of sight between each antenna and the region along the shower track where the radio emission takes place.
Note that variations of $n_i$ for different observer positions induce relative delays which ---while still sub-dominant compared to the general shape of the wavefront--- may become significant for the specific case of inclined showers, which is the focus of this work.

The curvature term $\mathcal{C}$ represents the spatial deviation from a plane wave\,:
\begin{align} \label{eq:wvf:curvature}
    c t_i -n_i \vec{k}\cdot\vec{x_i} = \mathcal{C} \ ,
\end{align}
This {\it wavefront curvature} will be  noted in the following $\mathcal{C} = \Delta t c/n$, $\Delta t$ being simply called the {\it time delay}.
\\

In the case of a hyperbolic curvature the wavefront times are given by~\cite{Corstanje_2015}
\begin{equation}
\label{eq:wvf:hyperbolic}
\frac{c}{n_i}\Delta t_i = c t_i -n_i \vec{k}\cdot\vec{x_i} = \sqrt{a^2+b^2r_i^2}-a
\end{equation}
where $r_i$ is the lateral distance from the antenna to the shower axis, and parameters $a$ and $b$ describe the hyperbolic shape of the wavefront. Parameter $a$ describes the curvature radius of the hyperbola at small values of $r_i$, while parameter $b$ gives its asymptotic slope at large values of $r_i$.

\begin{figure}[ht]
    \centering
    \includegraphics[width=0.99\linewidth]{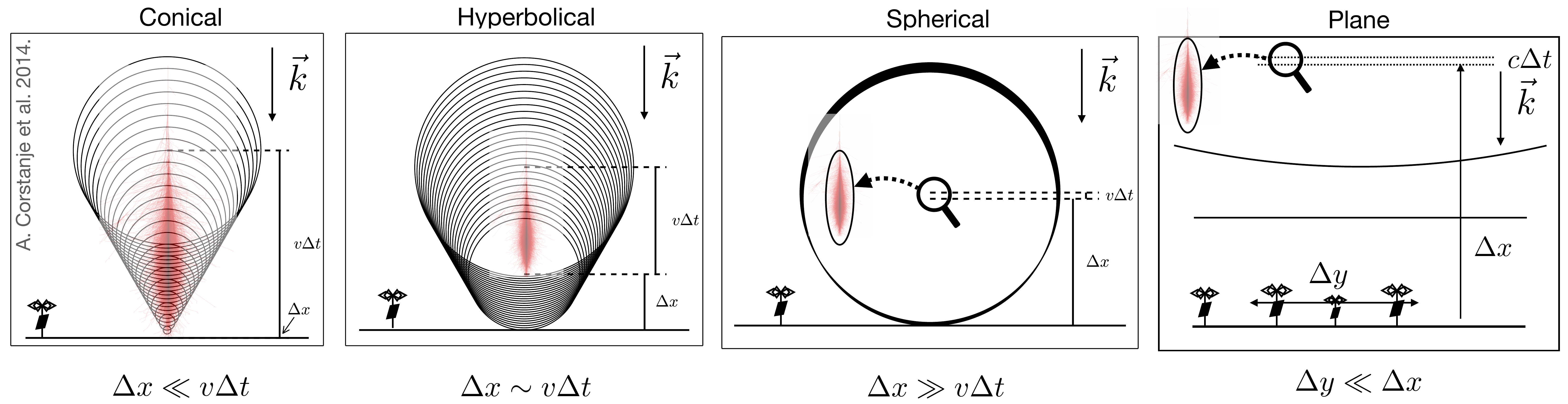}
    \caption{Wavefront models depending on the detector configuration, extension, and air-shower spatial extension. From {\it left} to {\it right}: conical: air-shower extension is much larger over the detector distance, hyperbolic: both air-shower extension and detector distance are comparable, spherical: detector distance is much larger than air-shower extension, and plane: the detector extension does not allow for a full sampling of the wavefront curvature. Adapted from~\cite{Corstanje_2015}.}
    \label{fig:wavefront_models}
\end{figure}

As mentioned in the introduction, the studies by LOPES and LOFAR were carried out with so-called {\it vertical} showers (zenith angle $\theta<60$\degree) and over detector arrays of limited size. In this specific case, the extension of the shower section around $X_{\rm max}$ where the bulk of electromagnetic emission takes place is comparable to its mean distance to an observer at ground, i.e., $\mathcal{O}$(10\,km). It is suggested in~\cite{Corstanje_2015} that this is the explanation for the hyperbolic shape of the wavefront, as illustrated in Figure~\ref{fig:wavefront_models}.

Yet inclined showers correspond to $X_{\rm max}$ located farther away from ground (larger than 100\,km for $\theta>80$\degree~\cite{Aab:2018ytv}) simply because of the larger atmospheric column density along the trajectory. This, together with simple geometric projection effects, also induces a very extended radio footprint at ground for inclined showers, with an electromagnetic pulse still measurable at lateral distances of several kilometers away from the shower axis. It is thus not excluded that the wavefront measured for these showers by setups such as GRAND~\cite{Alvarez-Muniz:2018bhp}, BEACON\,\cite{Wissel_2020} or AugerPrime~\cite{Castellina:2019irv} will differ from the ones observed by LOPES or LOFAR. The purpose of this paper is to study this specific issue and propose a dedicated method to reconstruct the wavefront of inclined showers, already briefly introduced in~\cite{Decoene:2021ncf}.

%
\subsection{Simulations}
\label{sec:study_method}

To perform this study, we use showers simulated with ZHAireS~\cite{2012APh....35..325A} version 1.0.28 for Aires version 19.04.00 \cite{Aires:url}, using the extended Linsley's atmospheric model and an exponential model for the index of refraction. The hadronic model used is Sibyll23c, the relative particle thinning is $10^{-5}$ with weight factor optimized for radio emission simulations and a time bin size of $0.5$\,ns. 

In our simulation, the magnetic field characteristics are computed with the IGRF13 model~\cite{igrf12} at a location set at $40.0\degree$ in latitude and $93.1\degree$ in longitude. The corresponding geomagnetic field inclination is $60.79\degree$, declination $0.36\degree$ and strength $55.997\mu$T. The ground altitude is set at $1086$\,m. The data set used in section \ref{wavefront_results} is composed of $3118$ simulations, with three distinct azimuth angles ($0, 90, 180\degree$), $13$ zenith angles distributed over logarithmic bins of $1/\cos{\theta}$ from $63$ to $87.1\degree$ and energies ranging from $0.02$ to $3.98$\,EeV in 25 logarithmic bins.

\begin{figure}
    \centering
    \includegraphics[width=0.8\linewidth]{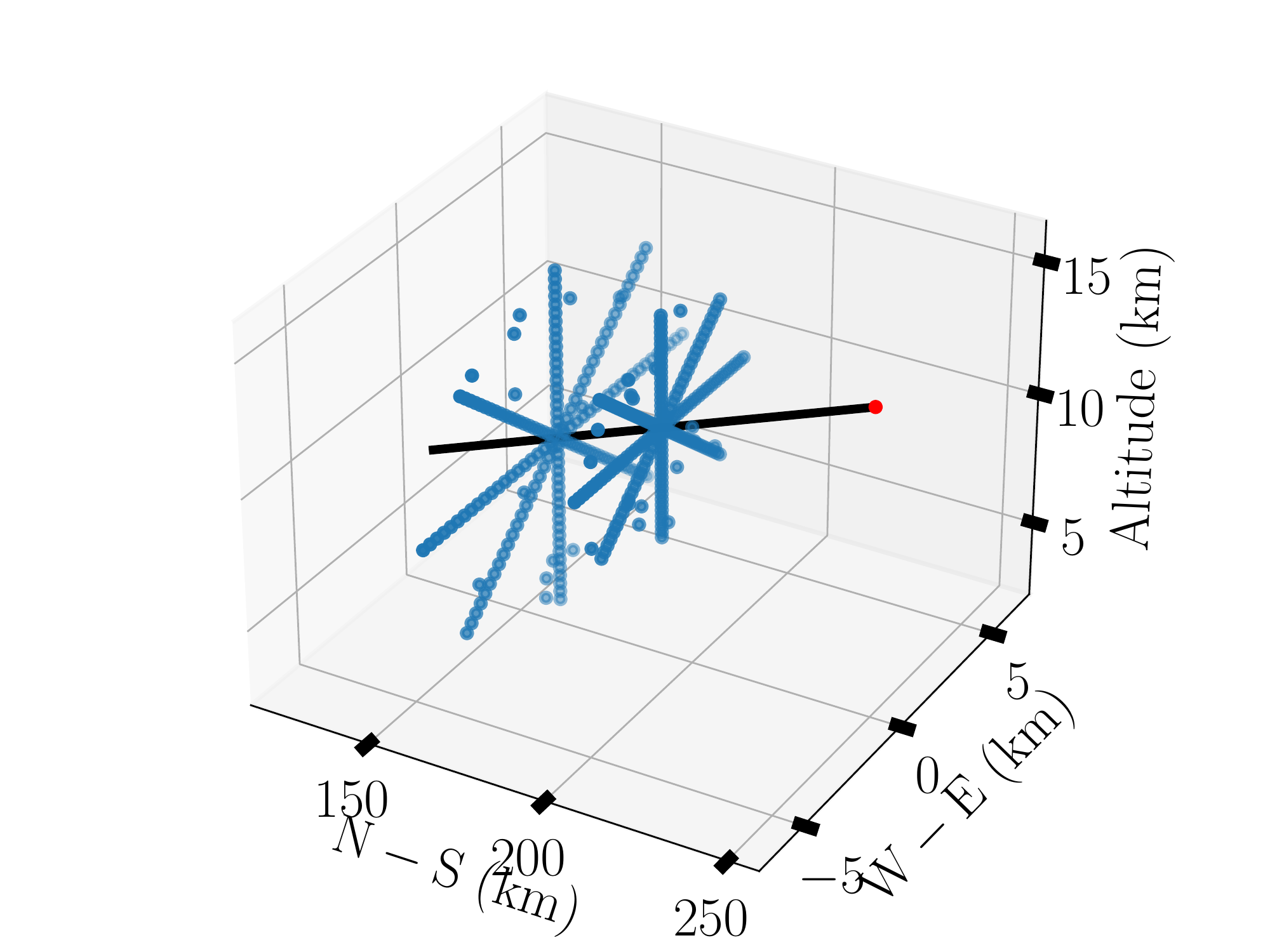}
    \caption{Example of a simulation layout. Antennas (blue dots) are placed along a star-shape pattern in two planes perpendicular to the shower axis (shown as a  black line) at specific distances from $X_{\rm max}$ (red dot). A few additional antennas are placed at random positions within the planes for cross-check.}
    \label{fig:starshape_config}
\end{figure}

These simulations are carried out over a 3D layout, composed of ten planes placed perpendicularly to the shower axis at longitudinal distances equal to 5, 7, 11, 17, 25, 38, 58, 88, 132 and 200\,km from $X_{\rm max}$. In each of these so-called {\it simulation planes}, 176 antennas are arranged in a star-shape pattern, as shown in Figure\,\ref{fig:starshape_config}. Each of the 8 arms is composed of 20 antennas and 16 additional antennas are randomly distributed within the starshape plane. The antennas on a given arm of the star-shape are separated by a constant angular step $\Delta \omega$, up to a maximal value $\omega_m$ = $4$\degree, where $\omega$ is the angular distance between the antenna and the shower axis as measured from $X_{\rm max}$.  

Unfiltered time-dependent components of the electric field along the South-North, West-East and vertical directions are computed at each antenna position. A mild cut on the electric field amplitude ---rejection of values lower than 22\,$\mu$V/m peak-to-peak--- is implemented in order to discard signals where the numerical noise may impact the determination of the trigger time.  

For selected signals, the Hilbert envelope of the electric field is then computed for each polarisation and the instant of the largest peak among the three polarisation is taken as the antenna trigger time.

\subsection{Treatment}
\label{wavefront_results}

\begin{figure}[ht]
    \centering
    \includegraphics[width=0.8\linewidth]{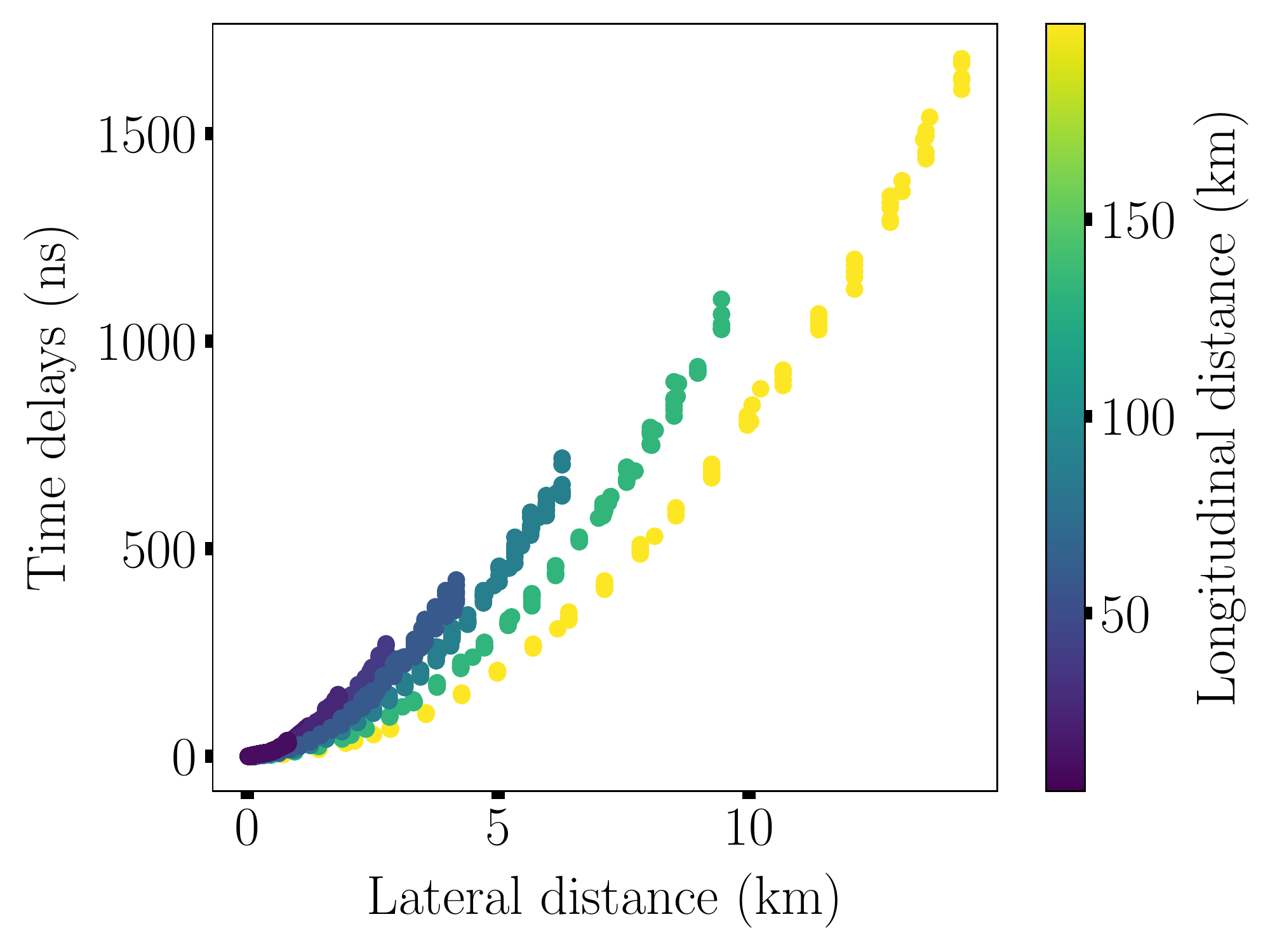}
    \caption{Time delays as a function of antenna distance to the shower axis for a shower induced by an iron primary of energy $E=3.98$\,EeV and zenith angle $\theta = 87.1\degree$ (i.e., down-going shower 2.9$\degree$ above horizon). Antennas from five simulation planes are shown here, corresponding to longitudinal distances of 38, 58, 88, 132, 200\,km from $X_{\rm max}$, as shown by the color code, to illustrate how the wavefront flattens with increasing distance to $X_{\rm max}$. The spread of the arrival time for antennas from a same plane and same lateral distance is generated by the asymmetry in the optical paths, an effect also pointed out in\,\cite{Decoene:2021ncf} and \cite{Schluter:2020tdz}. }
    \label{fig:extimedelays}
\end{figure}

We represent in Figure \ref{fig:extimedelays} the time delays (as defined in section \ref{sec:pheno}) for one shower from the simulation data set presented in section~\ref{sec:study_method}. The figure clearly shows that the wavefront deviates from a plane wave, with a curvature decreasing with longitudinal distance to $X_{\rm max}$. This shows that extended arrays, detecting radio footprints over tens of kilometers along the shower longitudinal axis, can be sensitive to the evolution of the wavefront. This suggests that this type of measurements may allow to  give an insight on shower development in a way similar to fluorescence detectors, a hypothesis which requires further investigation, but will not be discussed in more details here.

\begin{figure}[ht]
    \centering
    \includegraphics[width=0.49\linewidth]{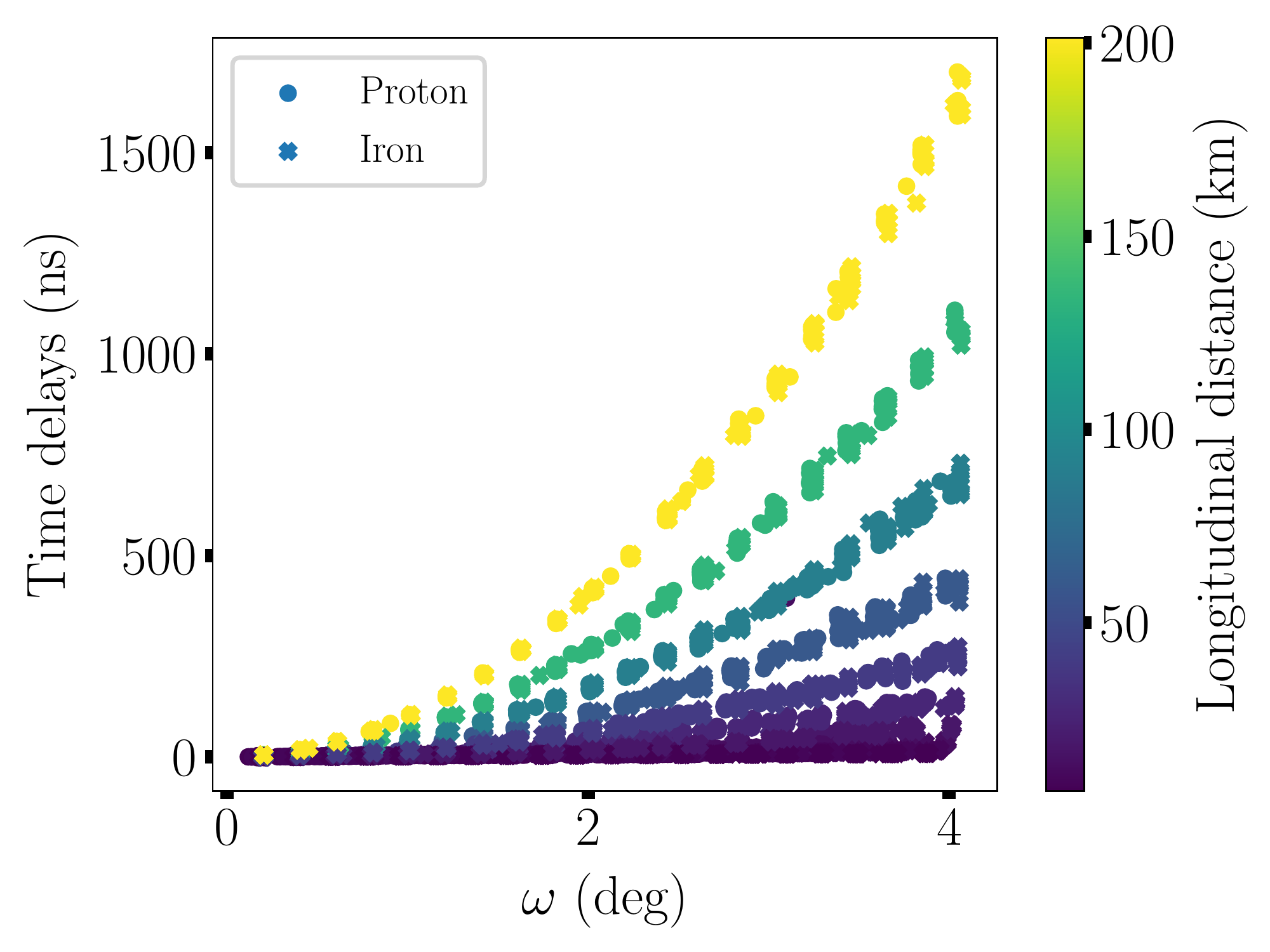}
    \includegraphics[width=0.49\linewidth]{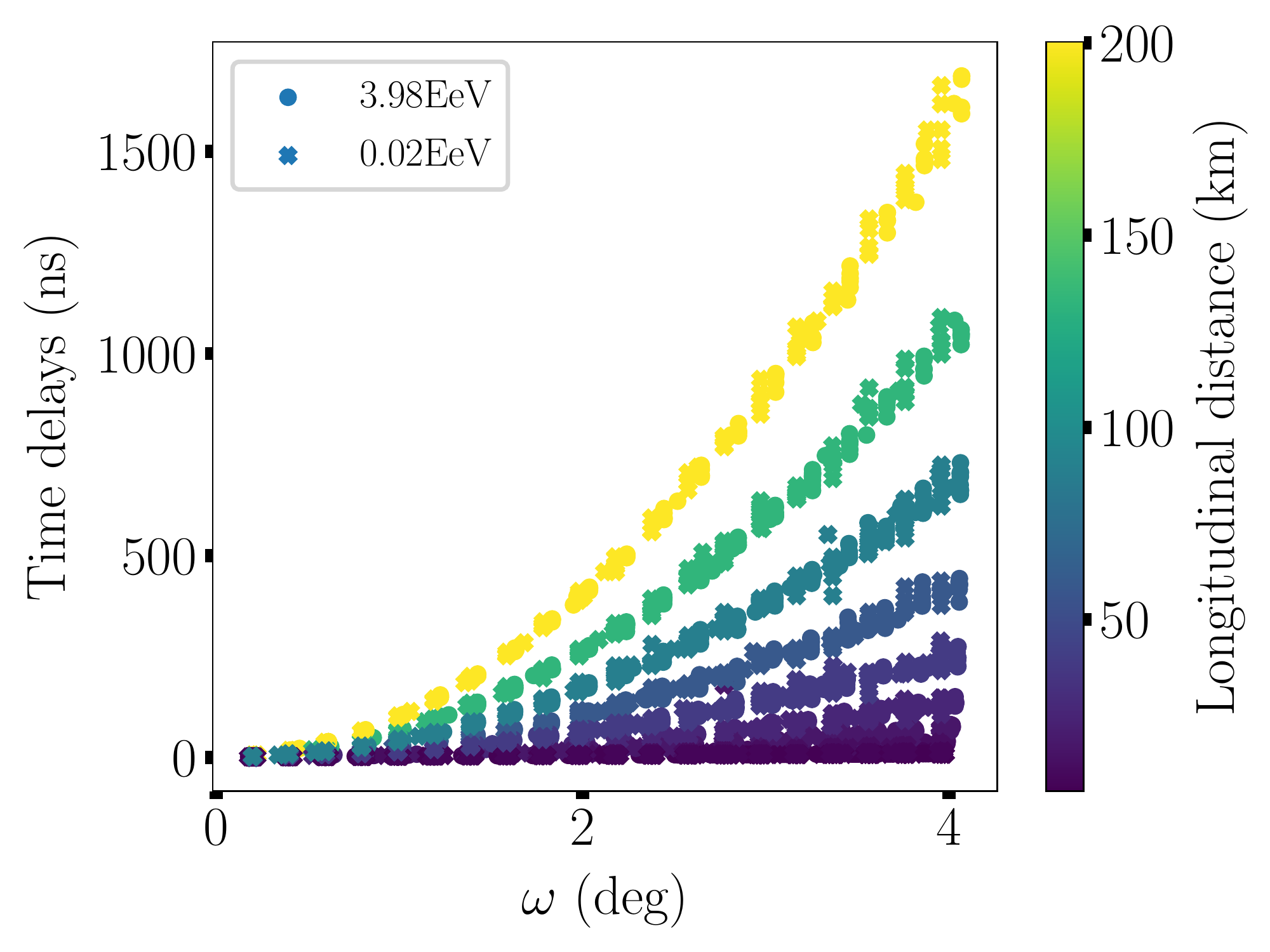}
    \\ \includegraphics[width=0.49\linewidth]{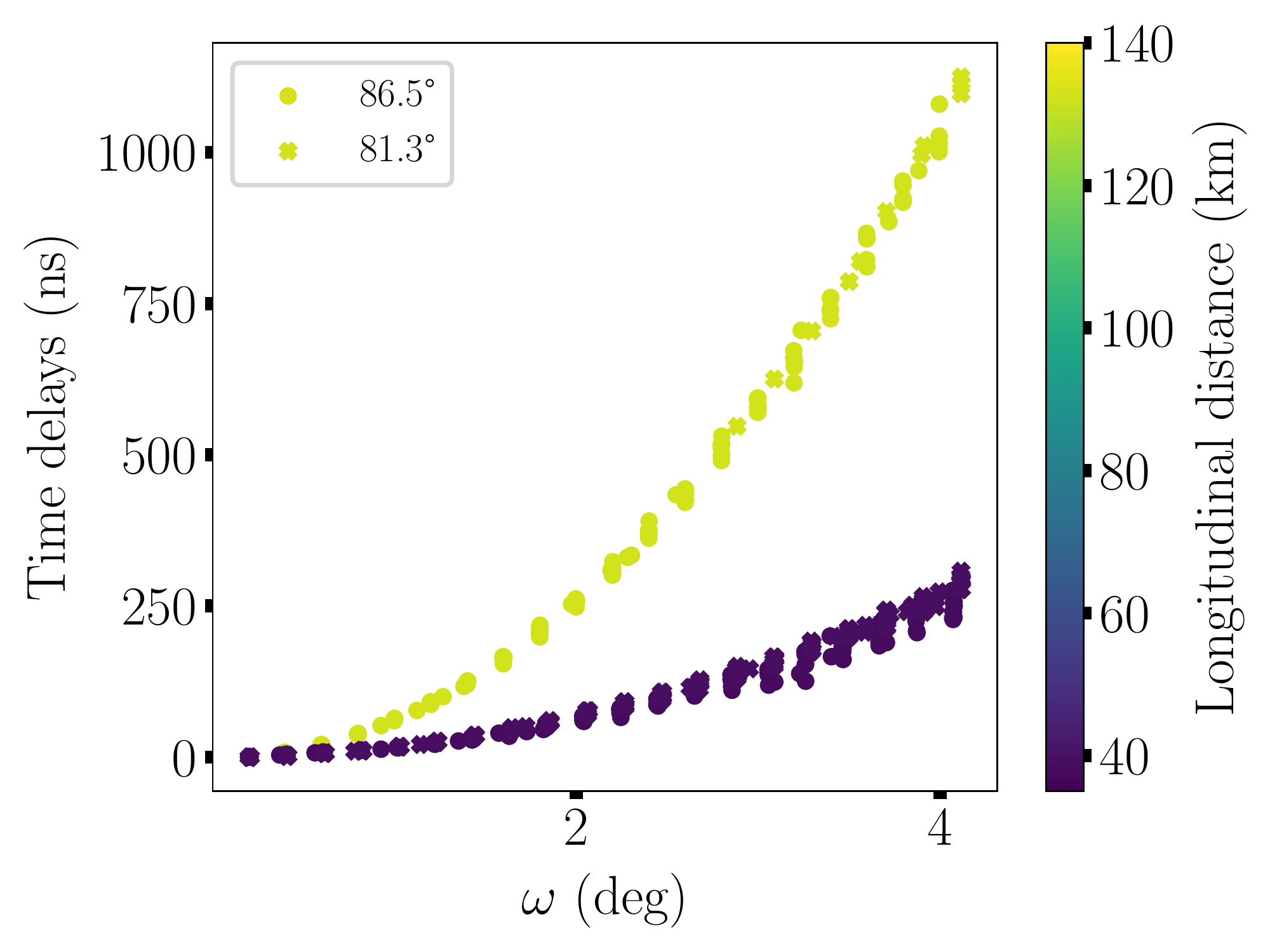}
    \includegraphics[width=0.49\linewidth]{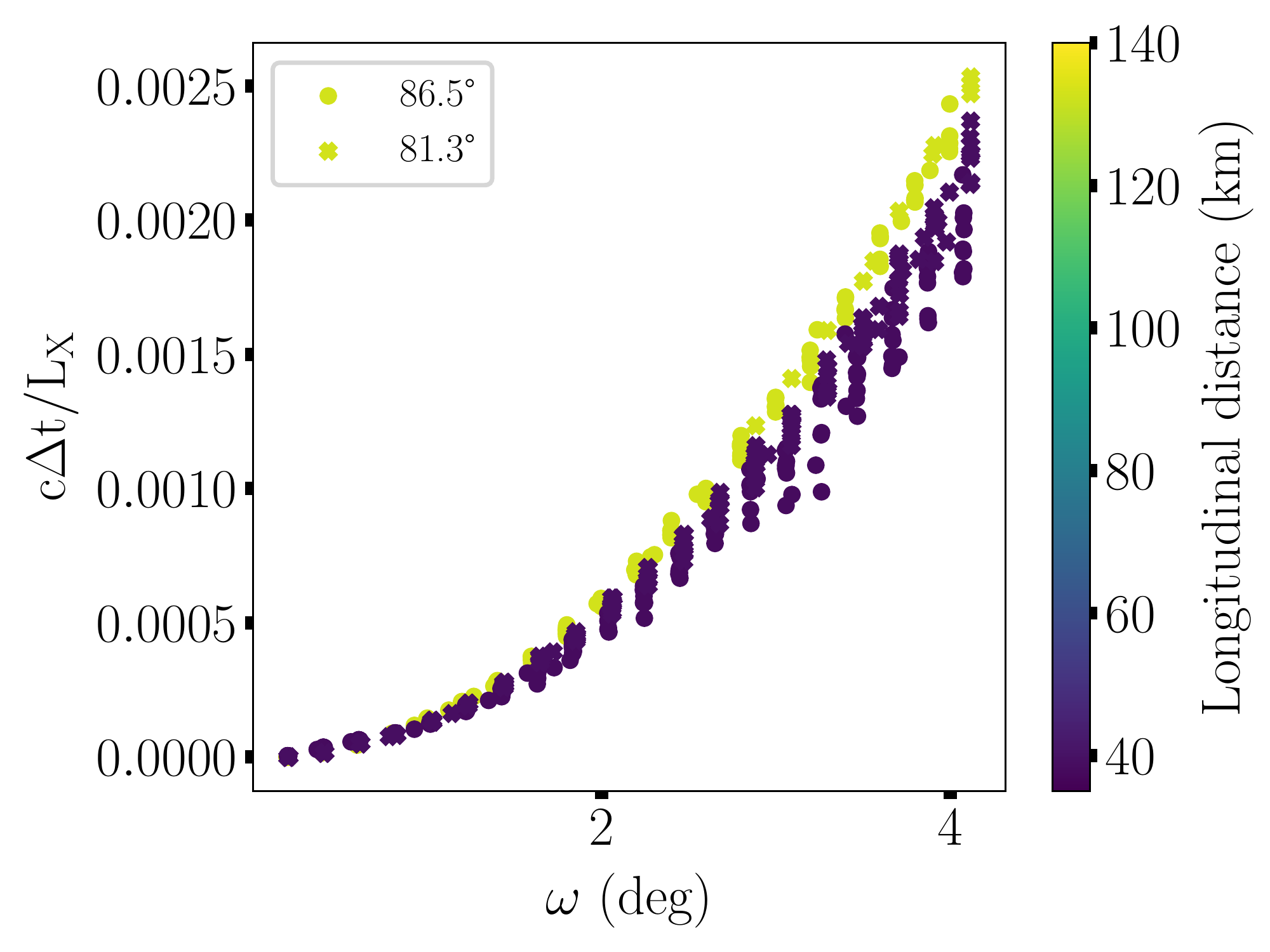}
    \caption{{\it Top-left:} Time delays as a function of angular distance to shower axis for antennas placed on the ten simulation planes considered in this study, identified through the color code associated with longitudinal distance. The data shown on this plot corresponds to two showers, with iron (x) and proton (o) primaries respectively, and otherwise identical parameters. {\it Top-right:} Same for two proton showers with energies of 3.98 and 0.02\,EeV, and otherwise identical parameters. {\it Bottom-left:} same for two different zenith values of $81.3\degree$ and $86.5\degree$, shown here for two simulation planes (38 and 133\,km) only. {\it Bottom-right:} wavefront curvature c$\Delta t$ normalized to $L_X$, the optical distance from antenna to $X_{\rm max}$ for the same data as in the left panel. Only slight differences between the two zenith angles can be observed for the 38\,km plane, but they remain below 5\,ns, showing that the shower front is not affected by zenith angle at first order.}
    \label{fig:extimedw}
\end{figure}

In Figure \ref{fig:extimedw}, time delays are represented for a few showers as a function of $\omega$, the angular distance from the antenna to the shower axis as measured from $X_{\rm max}$. These plots indicate that for a timing resolution $\delta t \lesssim$10 ns ---the nominal timing performance of the GRAND experiment~\cite{Alvarez-Muniz:2018bhp}--- the wavefront does not depend on energy, nature of the primary nor zenith angle. This is illustrated here for a few examples, but confirmed with a larger set of simulations. The dependency with distance to $X_{\rm max}$ shown in Figure \ref{fig:extimedelays} can therefore be considered as the only relevant parameter driving the wavefront shape at the level of precision achievable experimentally for timing. 

\begin{figure}[ht]
    \centering
    \includegraphics[width=0.99\linewidth]{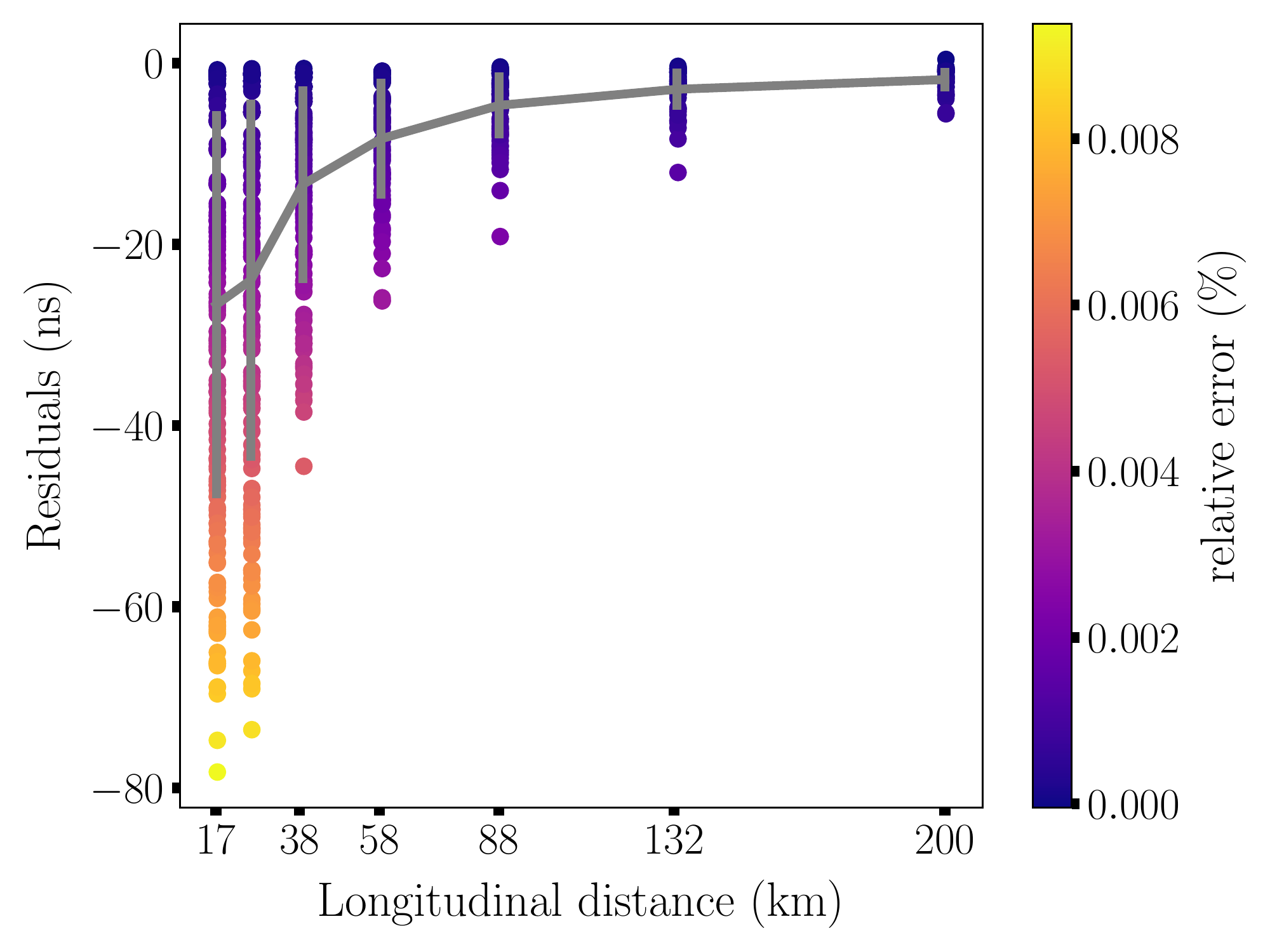}
    \caption{Residuals of a spherical fit applied to a shower induced by an iron of energy $3.98$\,EeV and zenith angle of $87.1\degree$. The shower is simulated over perpendicular antenna planes, and residuals are plotted against the planes longitudinal distance to $X_{\rm max}$. Each antenna in the simulation is represented as a dot which color corresponds to the residual normalised by the propagation time from the source to the antenna. 
    The mean value of the residuals, computed over each simulation plane, are also plotted in gray for each plane. It is lower than 10\,ns for longitudinal distances greater than 38\,km. Time reference is defined in this figure as the instant when the sphere crosses each simulation plane. Negative values thus show that the simulated wave front is not as curved as in a spherical model. Yet the relative error to the spherical model remains below 0.01\% for any plane, and residuals decrease with increasing longitudinal distances.} 
    \label{fig:residual}
\end{figure}

The fact that the wavefront shape only depends on the position of shower maximum advocates for a spherical shape of the shower wavefront. A spherical fit of the antenna trigger times was therefore performed, the PORT library~\cite{PORT:url} being used to minimize the quantity:
\begin{equation}
    R_{sph} = \frac{c}{n}\sum_{i=1}^{N}\lvert t_i-t_i^* \rvert
\label{fiteq}
\end{equation}
where $N$ is the number of selected antennas, $t_i$ the trigger time of antenna $i$ and $t_i^*$ its expected value for a spherical wavefront, given by:
\begin{equation}
    t_i^*= t_0+\frac{n\lvert\lvert \vec{x}_i-\vec{X}_{\rm e} \rvert\rvert}{c} 
\end{equation}
where $\vec{x_i}$ is the position of antenna $i$. The parameter $\vec{X}_{\rm e}$ can be considered in this model as the (static) point-like source of the radio emission. This parameter and the signal emission time $t_0$ are free parameters of the fit.
The fit is performed in all 3 dimensions, and the search is performed inside a cone section of parameters $\theta$, $\phi$ and $\rho$ to optimize the speed of the reconstruction process.  The parameters $\theta$ and $\phi$ cover a $2\degree$ range centered on values determined through an initial plane wave reconstruction, and the range for $\rho$ depends on a loose parametrisation.
A fit example shown in Figure~\ref{fig:residual} illustrates that the simulated wavefront is more curved that the result of the spherical fit, a result compatible with an hyperbolic hypothesis. Yet the effect decreases promptly with longitudinal distance, with a standard deviation of the fit residuals below 10\,ns for simulation planes beyond 38\,km.

\begin{figure}[ht]
    \centering
    \includegraphics[width=0.49\linewidth]{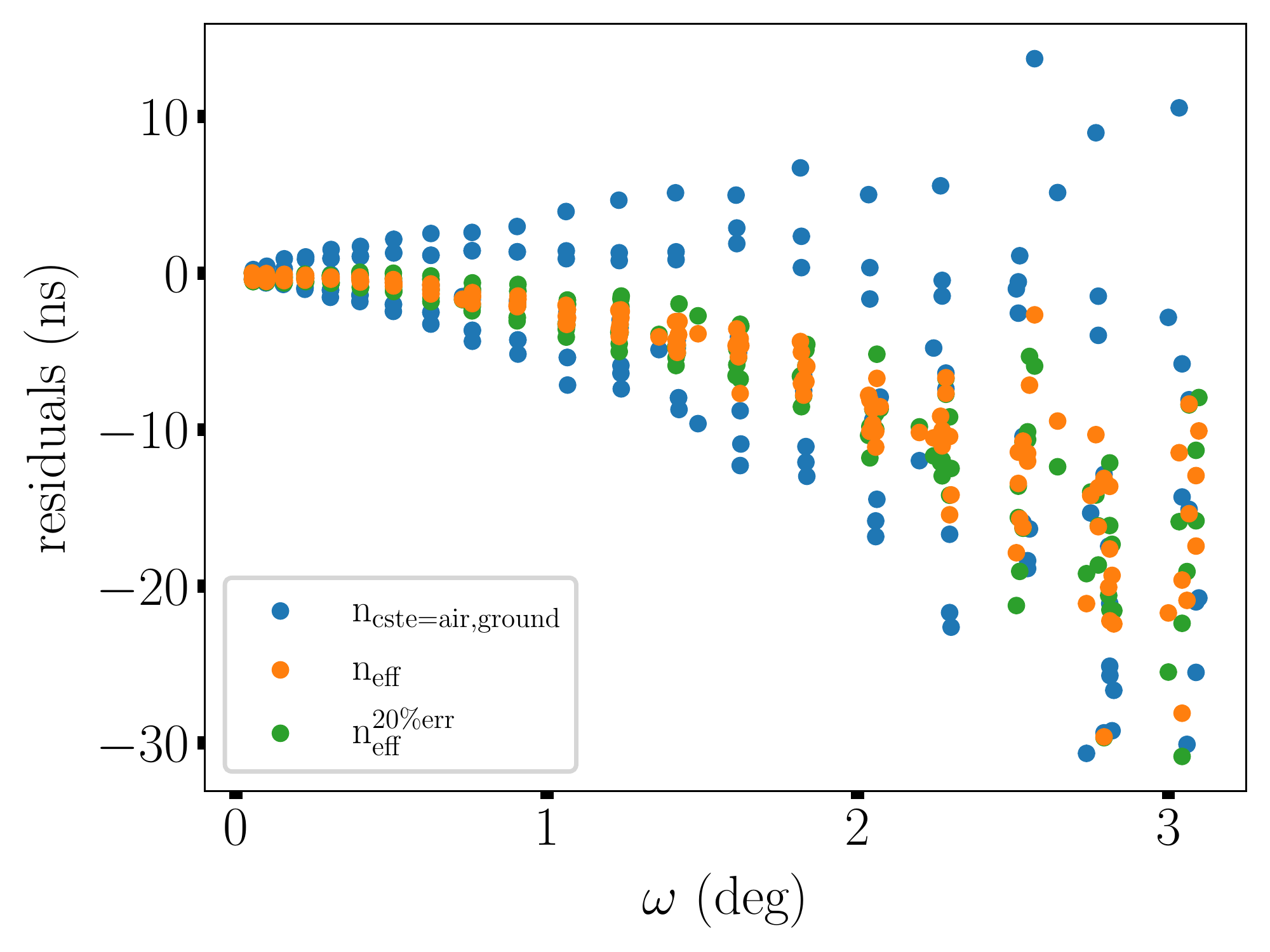}
    \includegraphics[width=0.49\linewidth]{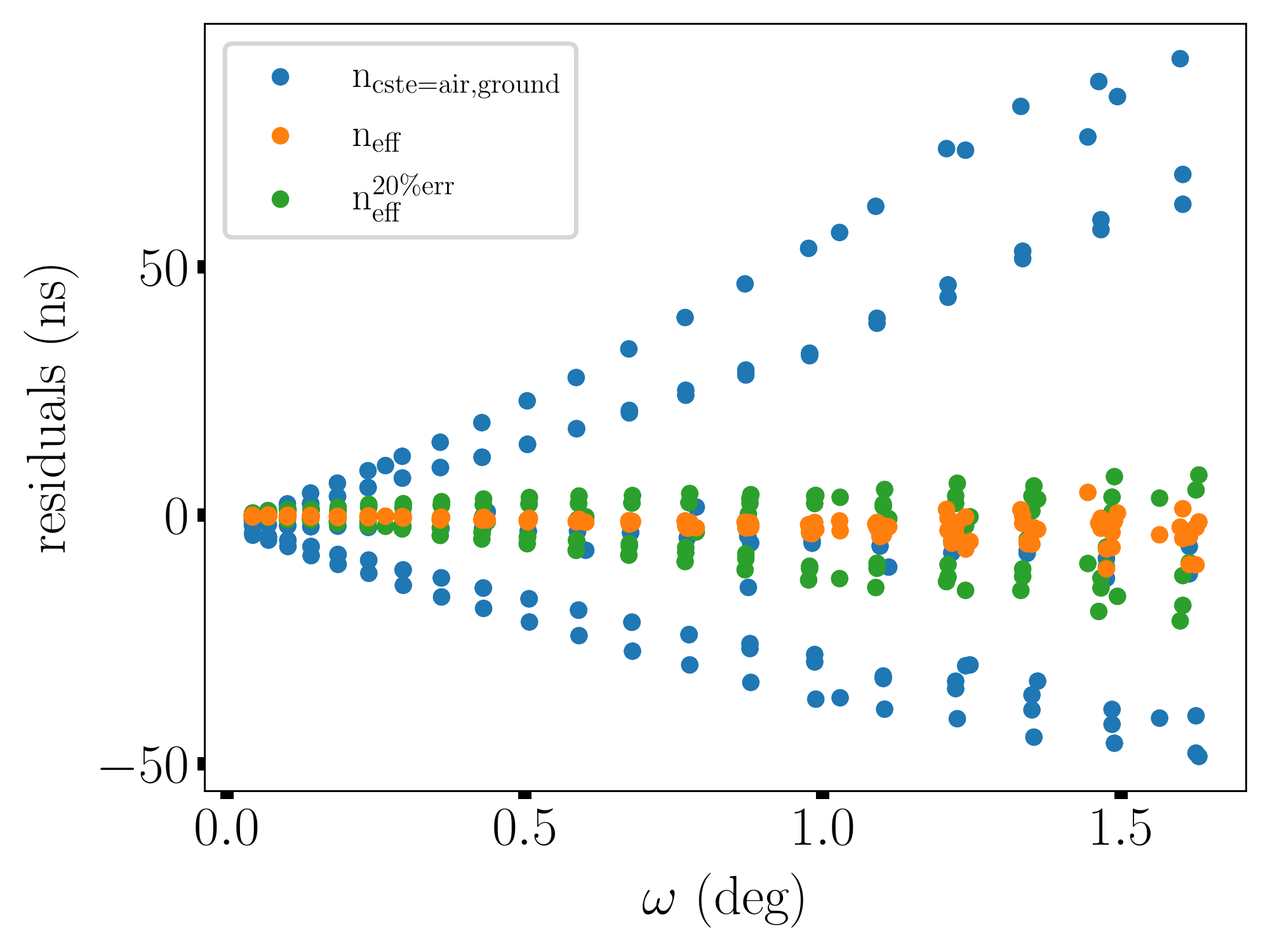}
    \caption{Residuals of a spherical fit for two simulated air-shower with iron primaries of energies $3.98$\,EeV. The signals are simulated for zenith angles of $81.3\degree$ ({\it left}) and $87.1\degree$ ({\it right}). Different values of refractive index $n$ are used in the fit equation: constant and equal to the ground level (blue), effectively integrated over the line of sight of each antenna (orange), and assuming a $20\,\%$ error (green).}
    \label{fig:residual2}
\end{figure}

Note that, since $n$ depends on the antenna position (i.e., the atmosphere is not isotropic), the proposed minimization results in a wavefront that is formally not a sphere nor a spheroid, but does correspond to a point-like source. We will however keep referring to it as a "spherical" wavefront for simplicity.
This distinction is however critical, since the radio emission of very inclined EAS propagates over very large distances ---tens to hundreds of kilometers--- before reaching the antenna, and consequently accumulates significant delays due to the refractive index changes. This effect is illustrated on Figure~\ref{fig:residual2}, where different values of the refractive index are used in equation \ref{fiteq}. This shows that a precise determination of the actual refractive index ---down to 20\%--- will be necessary in practice to allow for a proper description of the wavefront shape for the most inclined showers. This, and other experimental aspects such as optical refraction, effects of ground or clouds, will be studied in a future work.

\begin{figure}[ht]
    \centering
    \includegraphics[width=0.49\linewidth]{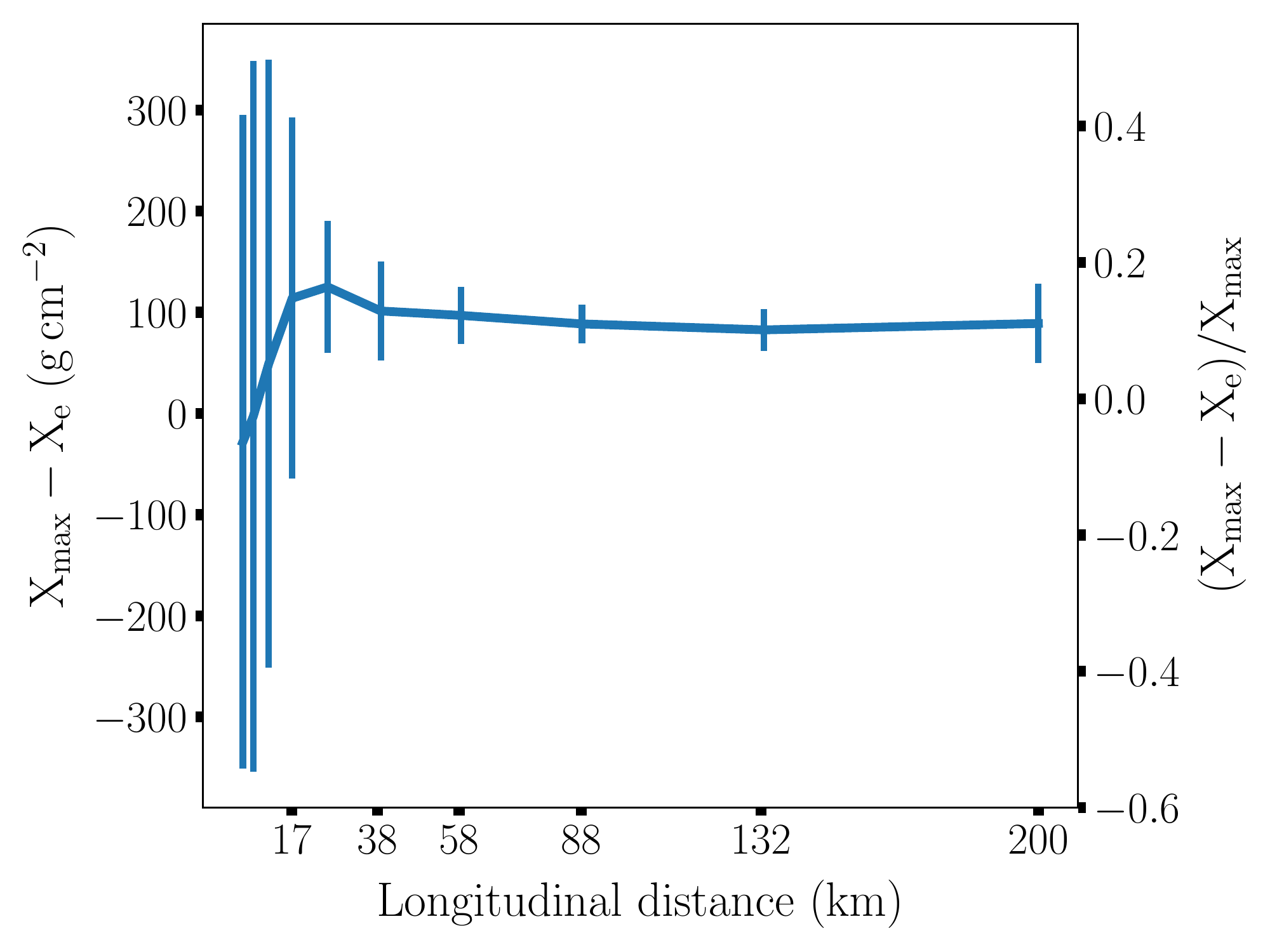}
    \includegraphics[width=0.49\linewidth]{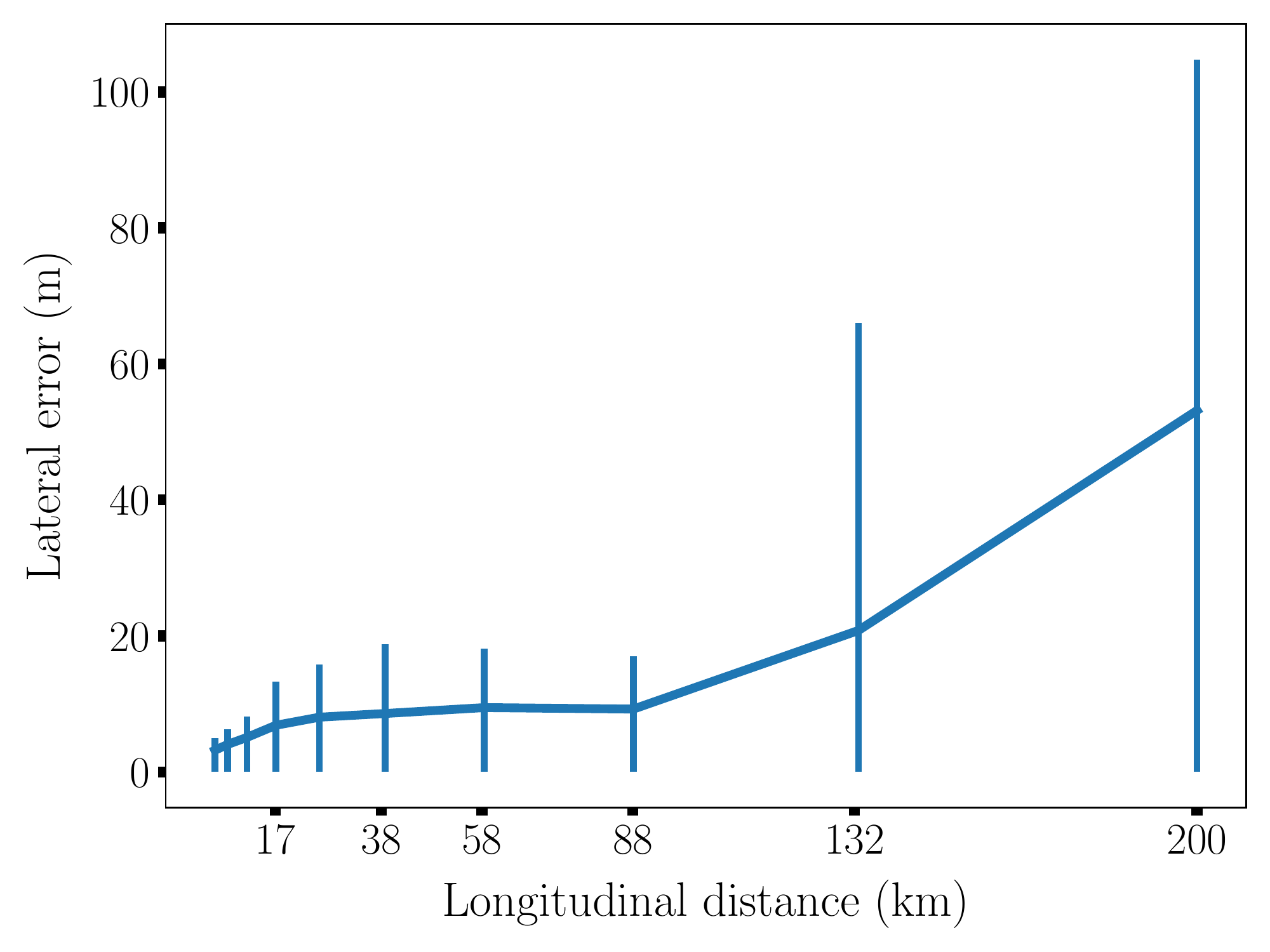}
    \caption{{\it Left:} Mean position of the radio emission point $\vec{X}_{\rm e}$ relative to the shower maximum for the ten simulation planes, represented here as a function of their longitudinal distance to $X_{\rm max}$. {\it Right:} same for the distance from emission point to shower axis, here in units of meters. In both cases error bars correspond to the standard deviation of each distribution. }
    \label{fig:errors_vs_planes}
\end{figure}

The spherical fit was applied independently to the ten simulation planes of each shower of the dataset presented in section~\ref{sec:study_method}. The distances between the reconstructed position of the source $\vec{X}_{\rm e}$ and the position of $X_{\rm max}$ are plotted in Figure~\ref{fig:errors_vs_planes}, together with the lateral distances from $\vec{X}_{\rm e}$ to the shower axis. The latter parameter is below 50\,m for any plane, an impressive result corresponding to a relative error better than 1\textperthousand. The average distance, along the shower axis, from $\vec{X}_{\rm e}$ to $X_{\rm max}$ position remains constant ---though not null--- within few percent for longitudinal distances larger than $17$\,km. This offset will be discussed in section \ref{sec:recons}. Here we will simply point that a clear drop in the average value of this parameter, together with an increasing dispersion, is observed for planes with longitudinal distance of 17\,km or below. We therefore consider that the spherical treatment is applicable only beyond this limit, a condition reached by showers with zenith angle $\theta \gtrsim 60\degree$ at energies around 10$^{17}$\,eV. 

This result can be understood as follows: a shower with inclination larger than $\sim$60$\degree$ has geometrical characteristics that can be associated with the situation depicted in the third panel of Figure \ref{fig:wavefront_models}: the extension of the shower section where the electromagnetic emission is sizeable is then negligible compared to its distance to ground, making a point-like hypothesis for shower radio emission valid to describe its wavefront.

\section{Reconstruction of the emission point}
\label{sec:recons}
In the previous section, we have seen that a spherical description of the EAS radio wavefront is adequate for showers with zenith angles larger than 60$\degree$. In a conservative approach, we will focus in the following on zenith angles $\theta \geq 71.6\degree$, and will study in more detail the performances of the spherical reconstruction and its implications in terms of emission point and physics behind it.
\subsection{Method}
We use in this section an antenna array deployed at ground, a more realistic case than the 3D pattern used in section \ref{sec:study}. We use a star-shape layout of 176 antennas, again with a constant step in $\omega$ and a maximal opening angle of $4\degree$ from $X_{\rm max}$. Reference \cite{Valentin_PhD} indicates that using hexagonal or rectangular antenna grids does not induce a significant degradation in the reconstruction performances, but this issue will be studied in more details in a later work.

20129 simulations were used for this study, with settings and parameters identical to those presented in section~\ref{sec:study_method} for energies, azimuth, and  zenith angles, except that zenith values $\theta < 71.6\degree$ are excluded, as mentionned already. The simulated traces are then filtered in the 30-80\,MHz frequency range, a typical band for EAS radio detection. Determination of the antenna trigger times from the simulated time traces is performed as in section \ref{sec:study_method}, but here only traces with peak amplitude of the Hilbert envelope beyond 45\,$\mu$V/m are selected. This corresponds to a signal-to-noise ratio (SNR) value of 3 where SNR is defined as ratio of peak-to-peak signal amplitude to $\sigma_{sky}$,  the average level of electromagnetic radiation induced by the Galaxy in the 30-80\,MHz frequency range. We find $\sigma_{sky}$ = 15$\mu$V/m, following the computation given in \cite{Decoene:2019izl}\,:
\begin{equation}
    {E_{\rm rms}}^2 = \frac{Z_0}{2}\int_{\nu_0}^{\nu_1}\int_{2\pi}B_{\nu}(\theta,\phi,\nu)\sin(\theta) d\theta d\phi d\nu
    \label{eq:skynoise}
\end{equation} 
where $B_{\nu}$ is the spectral radiance of the sky, computed with GSM \cite{Gal:2016} or equivalent codes, $Z_0=376.7$\,$\Omega$ the impedance of free space, and [$\nu_0,\nu_1$] the frequency range considered for detection.

We believe that an SNR value of 3 is an achievable trigger threshold in practice, yet for the purpose of completeness, a SNR value of 5 is also considered in the analysis, as will be detailed below.

\subsection{Peak time accuracy} \label{sec:peaktime_accuracy}
\begin{figure}[ht]
    \centering
    \includegraphics[width=0.49\linewidth]{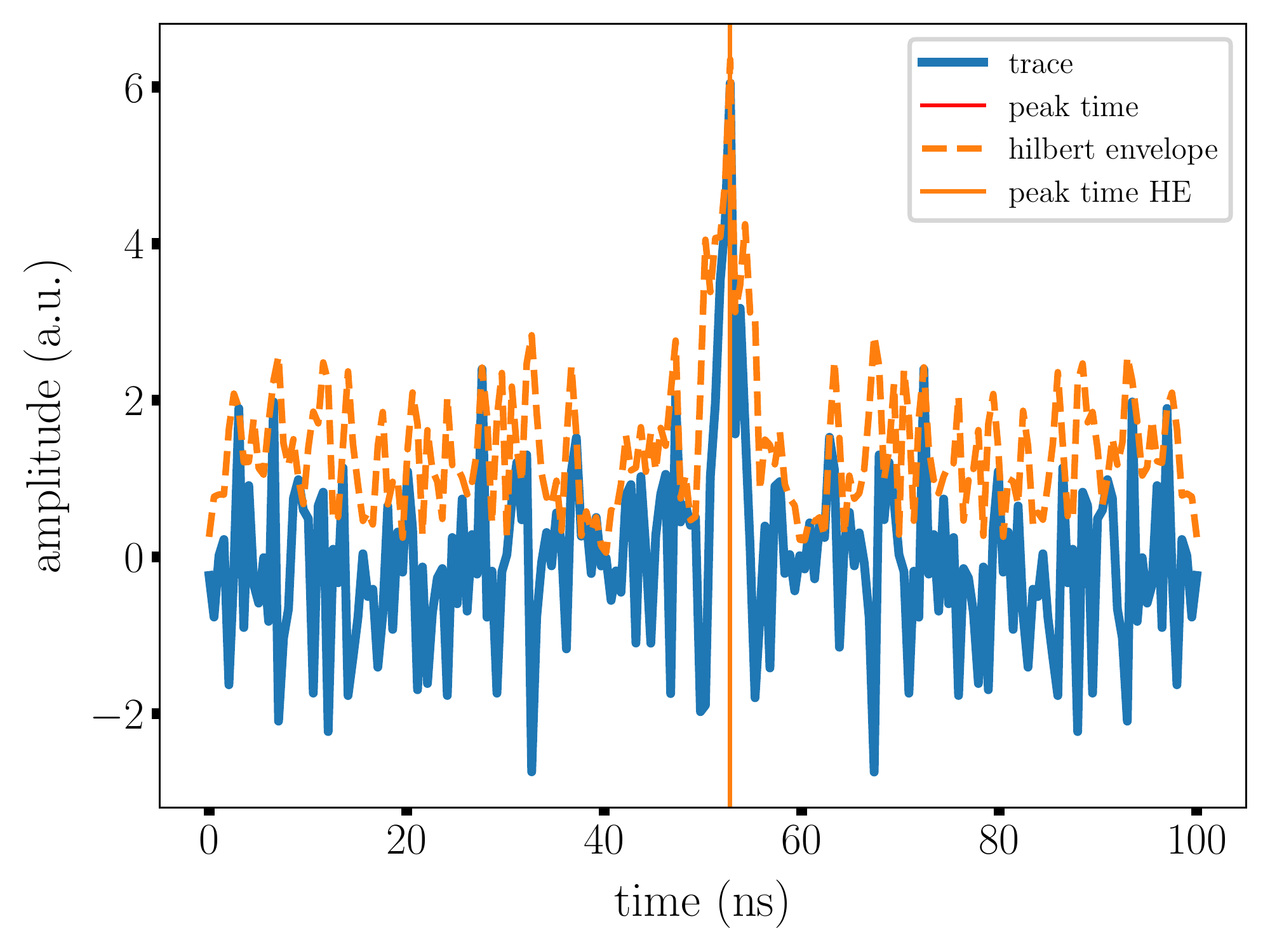}
    \includegraphics[width=0.49\linewidth]{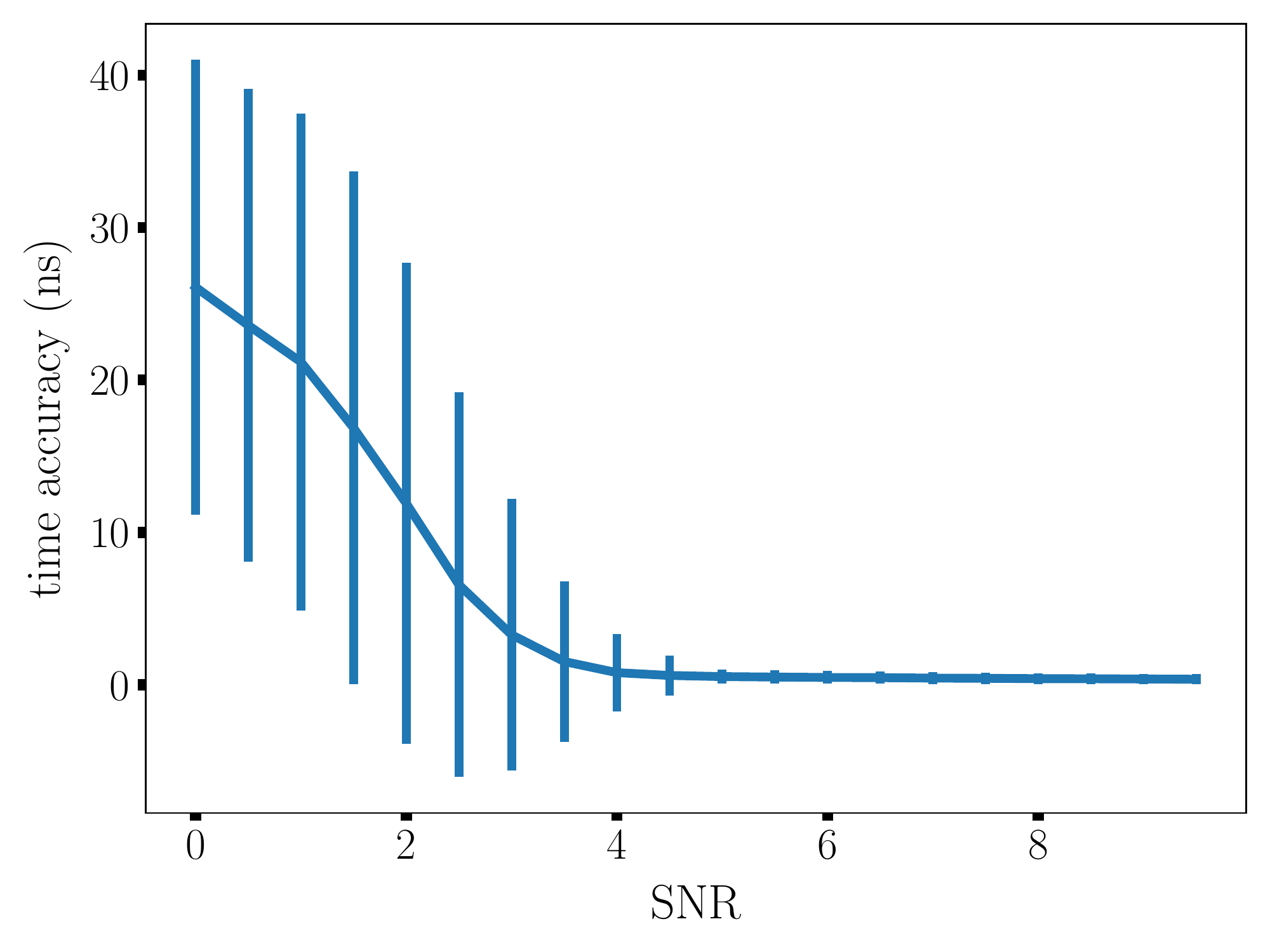}
    \caption{{\it Left:} Example of random gaussian noise, with a SNR of $4$  added to a simulated impulse trace (solid blue line). Its Hilbert envelope is shown in dotted orange, and position of the peak time marked as a vertical straight line in deep blue, while the true peak time of the impulse is represented as a vertical straight line in orange. {\it Right:} time accuracy as a function of SNR. Time accuracy is defined as the positive difference between the Hilbert envelope peak time and the true peak time. The values plotted correspond to the mean of the time accuracy distribution obtained for $10^5$ random traces, while the vertical bar correspond the standard deviation of this distribution. }
    \label{fig:time_SNR}
\end{figure}

In realistic conditions, the above-mentioned galactic emission and other background sources induce random noise in the recorded traces and thus impact the accuracy of the peak time determination.
In order to quantify this effect, we added random gaussian noise to simulated traces for SNR values ranging from 1 to 10, using the same SNR definition as above. For each value, a total of $10^5$ traces were generated, and for each trace the time difference between the true peak time and the peak time derived from the Hilbert envelope was computed. This \textit{toy model} study shows that the peak time accuracy is around $5$\,ns for SNR\,=\,3
(see Figure~\ref{fig:time_SNR}), a value smaller than the typical residuals of the spherical fit for longitudinal distances larger than our $17$\,km limit (see Fig. \ref{fig:residual}). Hence, we believe this effect should not affect significantly the reconstruction procedure, and it is thus not considered in this study. This also applies to the trigger time stamping by GPS, which resolution should also lie in the $5-10$\,ns in the case of GRAND. Yet again, both effects will be included in a latter work, which will present a practical case-study of the method detailed in this article.

\subsection{Results on radio source position} \label{sec:results_rec_ground}
\begin{figure}[ht]
    \centering
    \includegraphics[width=0.49\linewidth]{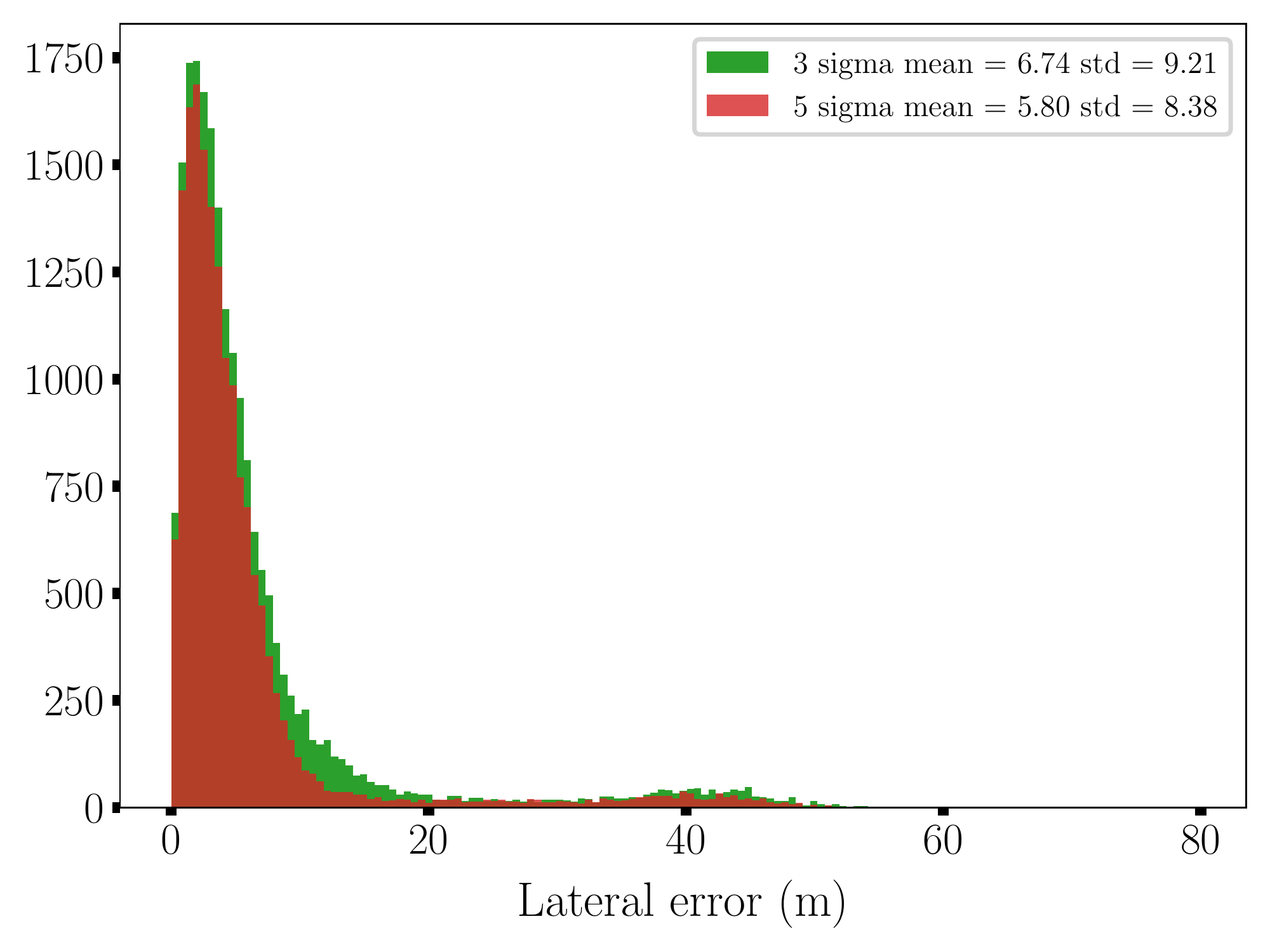}
    \includegraphics[width=0.49\linewidth]{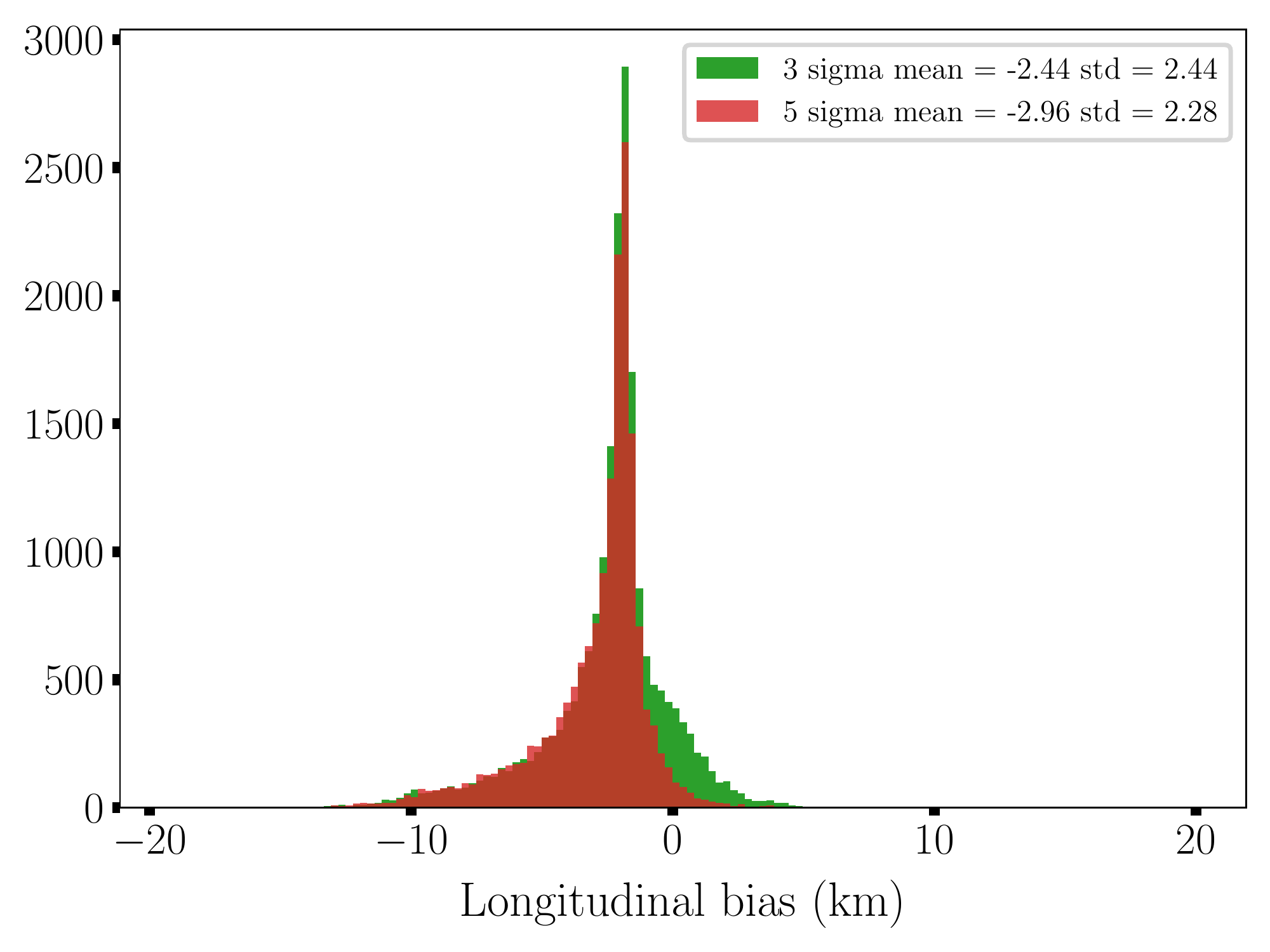}
    \caption{{\it Left:} Histogram of the lateral distance from the reconstructed position of the radio emission point to the shower axis for two SNR threshold values. {\it Right:} Histogram of the longitudinal bias (distance from the emission point to $X_{\rm max}$ projected along the shower axis). In both cases, both iron and proton primaries are considered.}
    \label{fig:recons_SNR}
\end{figure}

In Figure \ref{fig:recons_SNR} we plot the lateral distance of the reconstructed emission point  $\vec{X}_{\rm e}$ to the shower axis for SNR threshold values of 3 and 5. The large extension of the radio footprint at ground, with signal at lateral distances larger than 1\,km from the shower axis, provides a powerful lever arm to constrain the spherical wavefront adjustment, and allows to achieve a 10\,m rms accuracy on this parameter.  

\begin{figure}[ht]
    \centering
    \includegraphics[width=0.49\linewidth]{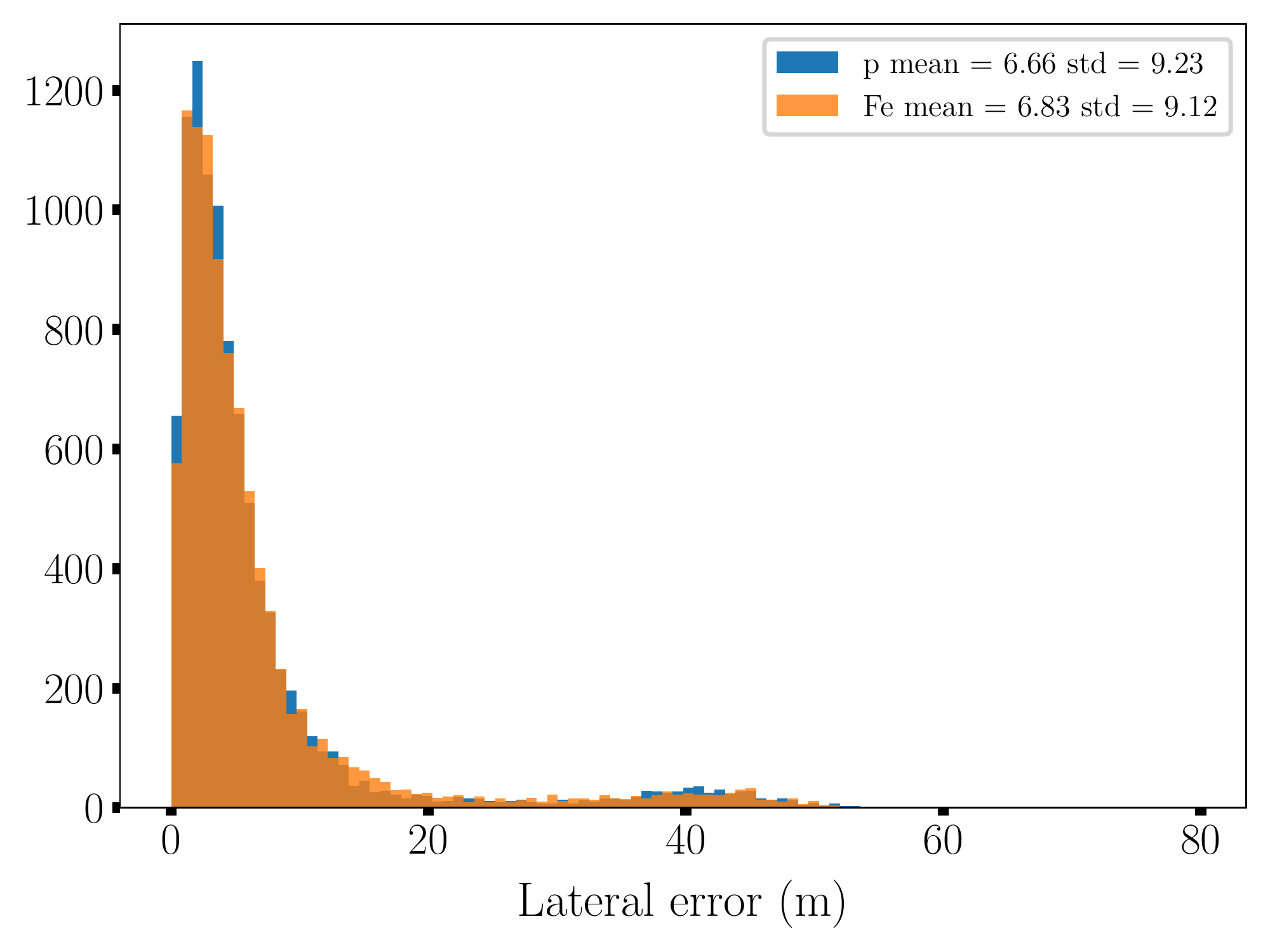}
    \includegraphics[width=0.49\linewidth]{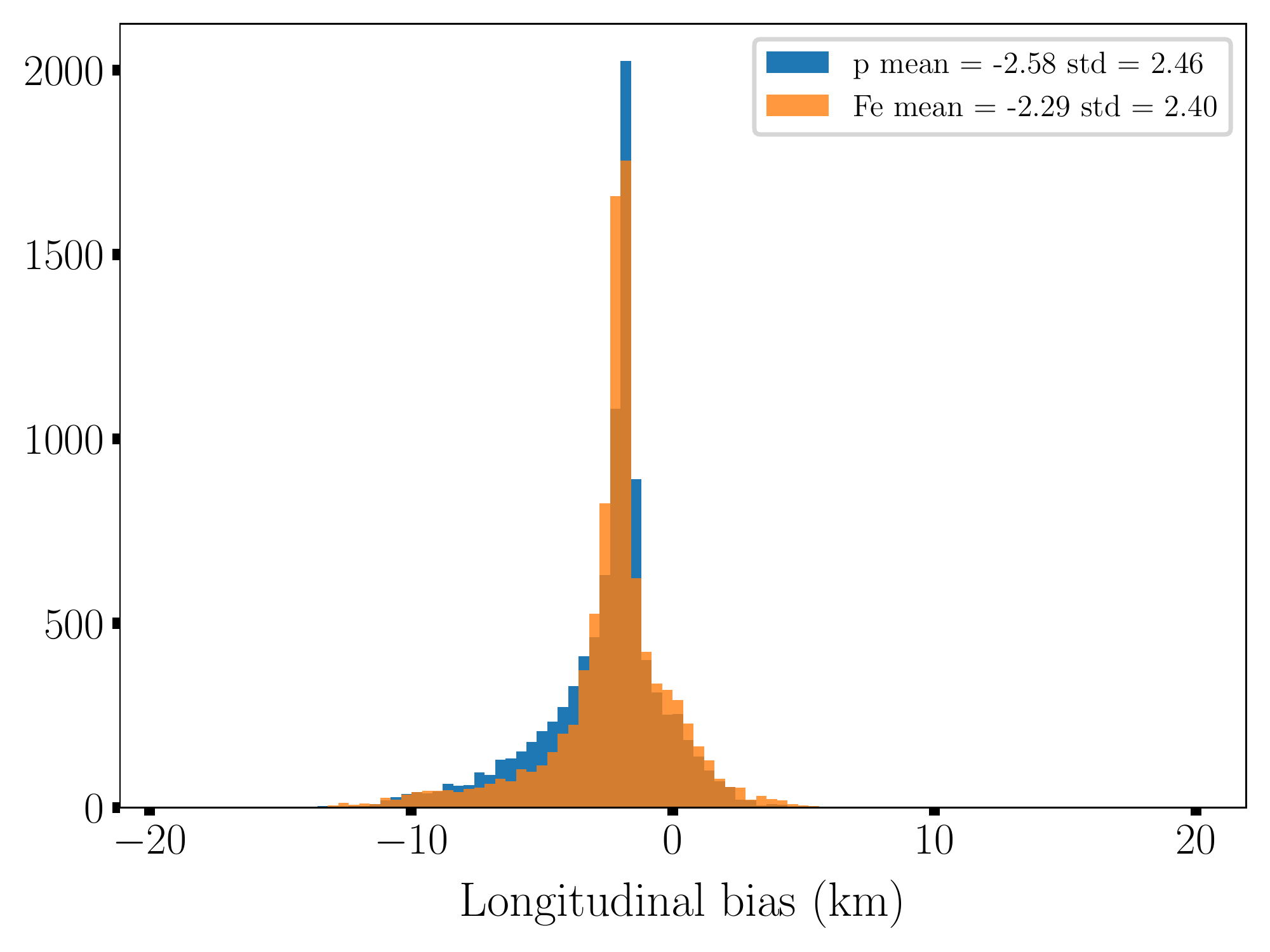}
    \caption{Same as Figure~\ref{fig:recons_SNR} for a SNR threshold value of $3$, and for protons and irons.}
    \label{fig:recons_p_Fe}
\end{figure}

\begin{figure}[ht]
    \centering
    \includegraphics[width=0.49\linewidth]{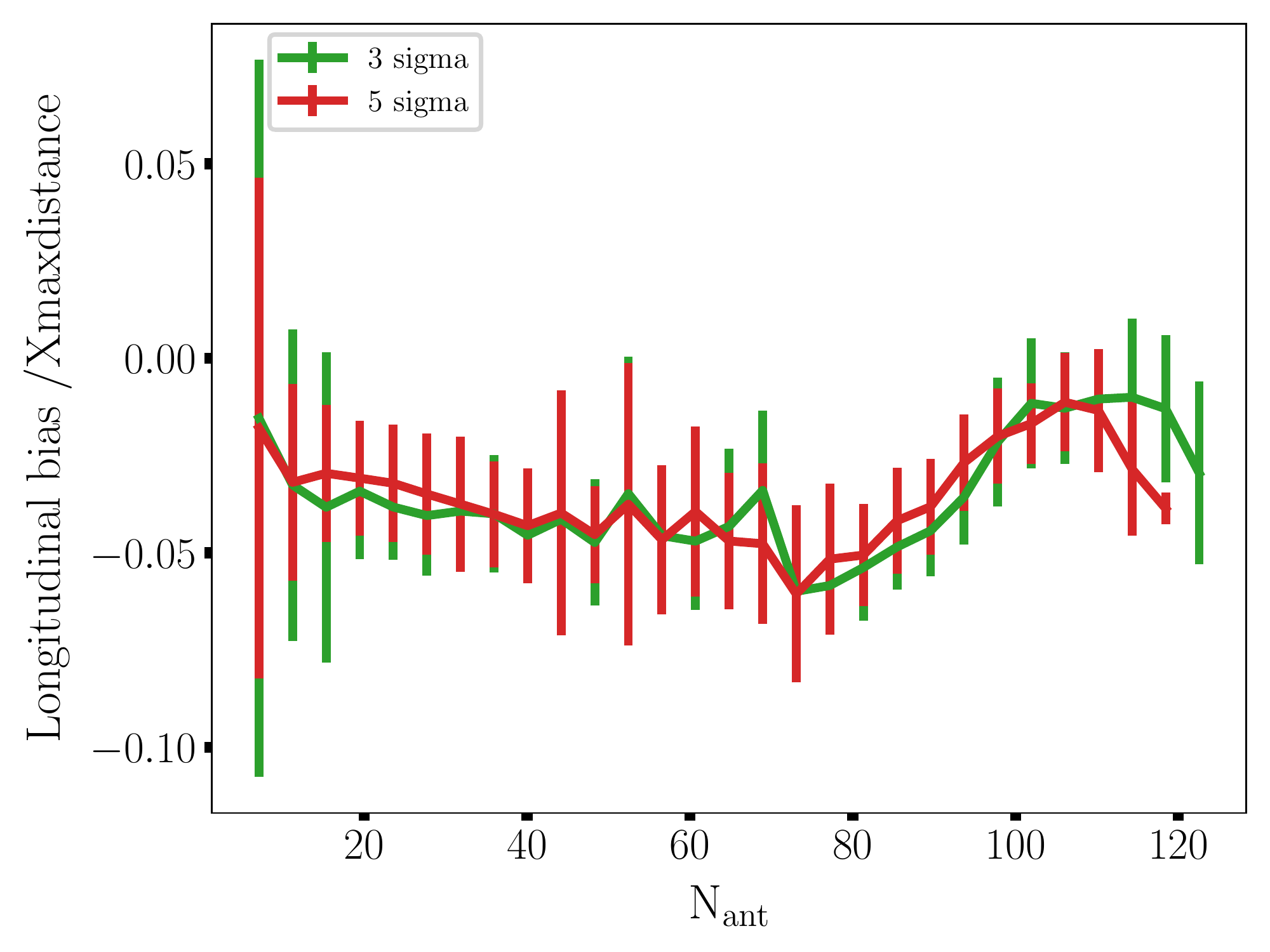}
    \includegraphics[width=0.49\linewidth]{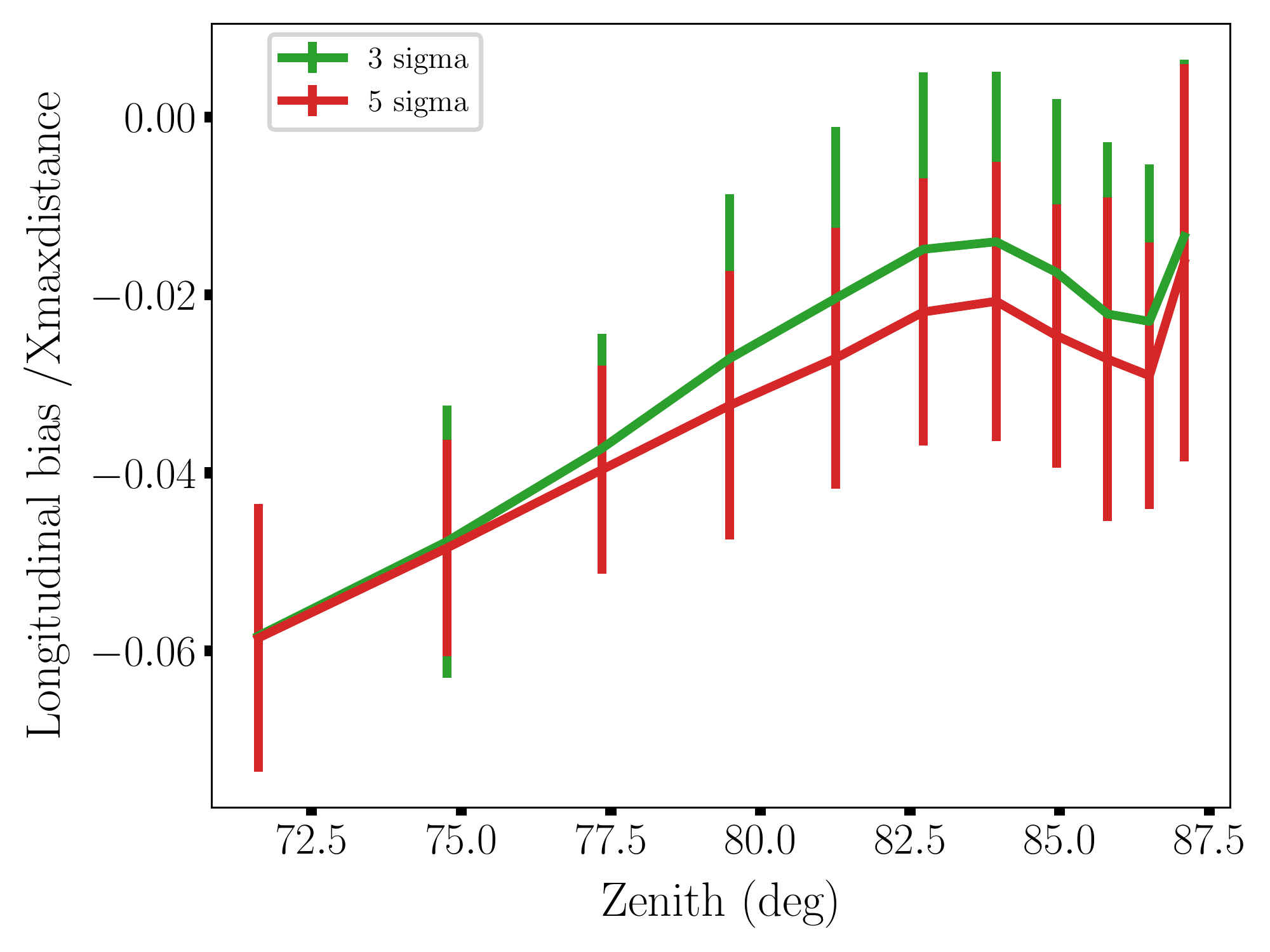}
    \caption{Evolution of the longitudinal bias ratio over the total $X_{\rm max}$ distance as a function of the number of antennas in each footprint ({\it left}), and the zenith angle of the air-shower ({\it right}), for both iron and proton primaries. }
    \label{fig:longbias_vs}
\end{figure}

We also compare in Figure \ref{fig:recons_SNR} the position of $\vec{X}_{\rm e}$ projected along the shower axis relative to the position of $X_{\rm max}$ for SNR values of 3 and 5. This so-called {\it longitudinal bias} exhibits a much larger rms value (2.4\,km both for proton and iron primaries) than the lateral distribution. This can be explained by straight-forward geometric arguments, i.e., at large radii a spherical fit allows for a much weaker constraint on the longitudinal position of the point source compared to its lateral distance to the shower axis. Yet it can be noted that the center of the spherical fit --- the {\it effective} source of the radio emission--- is  systematically reconstructed at a position higher in atmosphere than the maximum of shower development. The bias value is small compared to the total distance between the shower core and $X_{\rm max}$ (typically below $\sim5\%$, see Figure \ref{fig:longbias_vs}),  and does not significantly depend on threshold (Figure \ref{fig:recons_SNR}), nature of the primary (Figure \ref{fig:recons_p_Fe}), zenith nor number of antennas in the events (Figure \ref{fig:longbias_vs}). Yet it decreases with energy, as will be seen in Figure \ref{fig:radiogram_mean}.

We detail below reasons why the maximum power of the emission could occur close to $X_{\rm max}$ where the charged particle number is the highest. The geomagnetic contribution to the electric radiation by the shower is  directly related to the derivative of the number of particles~\cite{Scholten:2007ky}, a term maximal before $X_{\rm max}$ position. We can briefly detail this as follows\,: the electric field can be derived from the  Li\'enard-Wiechert vector potential $\vec{A}\qty(t, \vec{x})$, for an observer at time $t$ and location $\vec{x}$. Neglecting dipole radiation (subdominant with respect to the other components), the derivation can be simply reduced to:

\begin{align}
    \vec{E}\qty(t, \vec{x}) \sim -\partial_{t} \vec{A}\qty(t, \vec{x}) \ .
\end{align}

In~\cite{Scholten:2007ky}, the authors show that $\vec{A}\qty(t, \vec{x}) \propto \vec{j}\qty(t, \vec{x})$ for EAS, where $\vec{j}\qty(t, \vec{x})$ is the charge current induced from the distribution of electrons and positrons $N_e\qty(t, \vec{x})$ inside the shower pancake, hence $\vec{j}\qty(t, \vec{x}) \propto N_e\qty(t, \vec{x})$. 
Therefore, the maximum of the emission is reached when the convolution between the particle distribution and its derivative tends to its maximum. This point is expected to be located before the particle maximum $X_{\rm max}$, given the typical particle distributions observed in EAS. This is also in agreement with results presented in e.g.,~\cite{schoorlemmer2020radio} where the maximum of the signal power appears to peak before the $X_{\rm max}$ location.

Yet our study is based on wavefront times rather than signal power, hence this argument has to be considered cautiously. The link between the wavefront trigger times and a specific location in the shower is difficult to derive since the optical paths are not constant and equal between different observers and different regions of the shower. This implies that a signal produced later in time may arrive earlier to a given observer position, and consequently blur the tracking of a static emission point into a moving emission region. 
Even more important, it should be kept in mind that our ad hoc point-source model is not related to the physics of the emission, but to the observation that inclined showers present a nearly spherical wavefront. 

To conclude this paragraph, we will point again that proton and iron primaries exhibit similar mean values for the longitudinal bias (-2.58 and -2.29\,km respectively, see Figure \ref{fig:recons_p_Fe}). This hints towards the idea that the offset observed in $X_{\rm max}$ for these two primaries also exists for $\vec{X}_{e}$, which may thus correlate to the nature of the shower primary.
This is detailed in the following section.


\subsection{Radio emission point position as a new proxy for mass identification?}
For each shower, we computed the so-called {\it radio grammage}, the atmospheric column density along the true shower direction from its entrance in the atmosphere to the position of the radio emission point $\vec{X}_{e}$ reconstructed through the spherical fit detailed above. 

\begin{figure}[ht]
    \centering
    \includegraphics[width=0.99\linewidth]{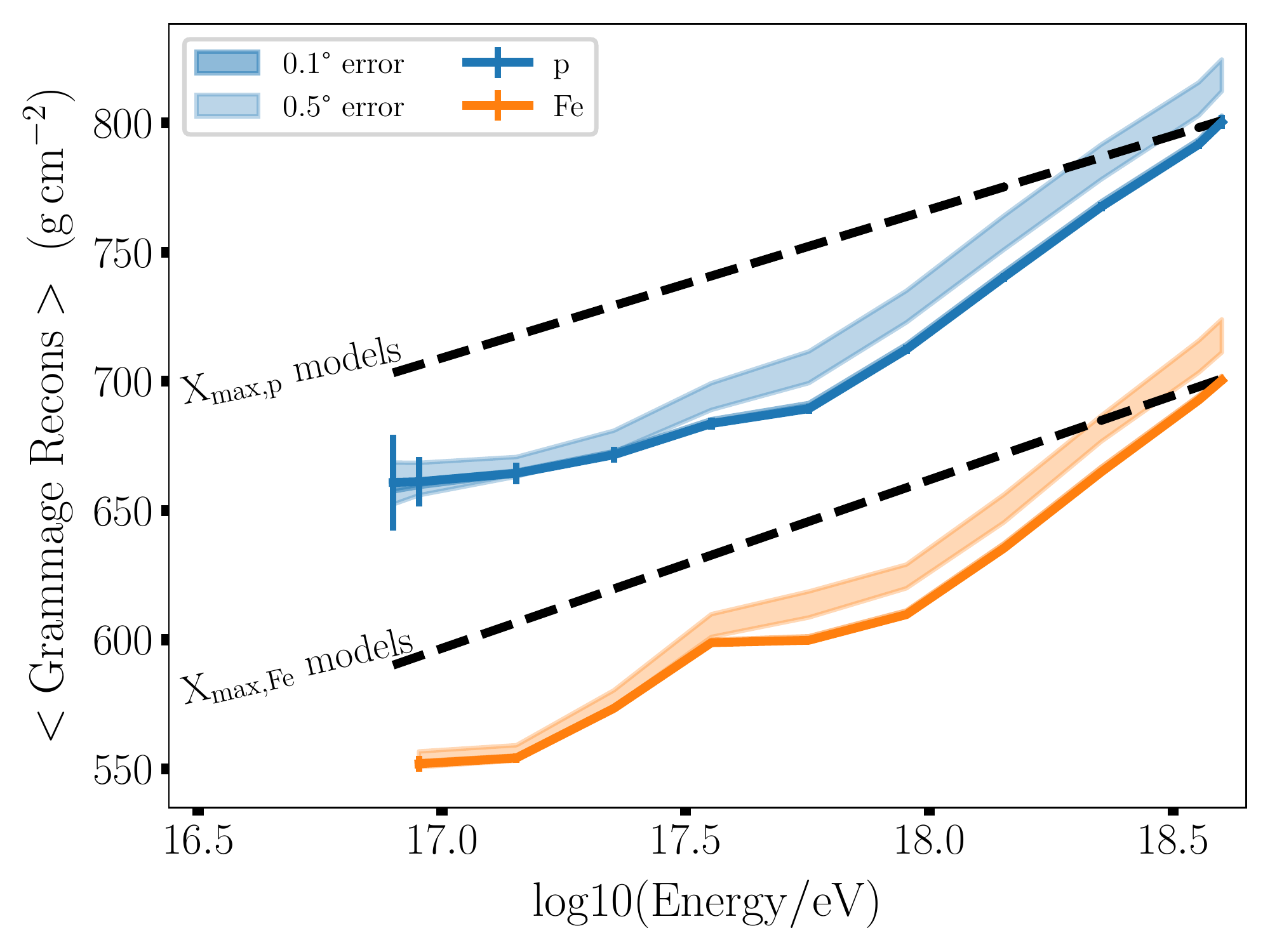}
    \caption{Mean value of the radio grammage distribution per energy slices for Proton (blue line) and Iron (orange line). The error-bars show the statistical fluctuations, taken equal to $\sigma_{X_{\rm e}}/ \sqrt{N}$ where N is the number of simulated showers per energy slice. The shaded areas correspond to the additional uncertainties associated to error on the direction of origin of the showers for 1-$\sigma$ values of 0.1{\degree} (dark) and 0.5{\degree} (light). See text for detail. The black dotted lines correspond to the standard $X_{\rm max}$ elongation rates for proton and iron primaries. It should be stressed here that even if significant on the average, the different values of radio grammage for iron and proton primaries do not allow for a shower-to-shower identification and clear determination of the primary cosmic ray nature, given the large fluctuations expected for the radio grammage variable, as displayed in Figure  \ref{fig:radiogram_std}. }
    \label{fig:radiogram_mean}
\end{figure}

We represent in Figure \ref{fig:radiogram_mean} the mean values of the radio grammage as a function of energy for showers induced by proton and iron primaries separately. It appears from this plot that the radio grammage follows a trend comparable to what is observed for (standard) $X_{\rm max}$ elongation rate, with deeper showers with increasing energies. 
The shallower position of $\vec{X}_{\rm e}$ with respect to $X_{\rm max}$ observed over most of the energy band considered here (see section \ref{sec:results_rec_ground}) also translates in a radio grammage offset around 50\,g\,cm$^{-2}$ at 10$^{17}$\,eV. Yet the somewhat steeper slope of radio grammage elongation rate ($\sim$100-120\,g\,cm$^{-2}$/decade) at energies above $10^{18}$\,eV leads to similar values for $\expval{X_{\rm e}}$
and $\expval{X_{\rm max}}$ at 10$^{18.5}$\,eV. We will refrain from providing an interpretation for this effect, since our ad hoc point-source model does not hol on physical basis, as discussed in the previous paragraph. 

\begin{figure}[ht]
    \centering
    \includegraphics[width=0.99\linewidth]{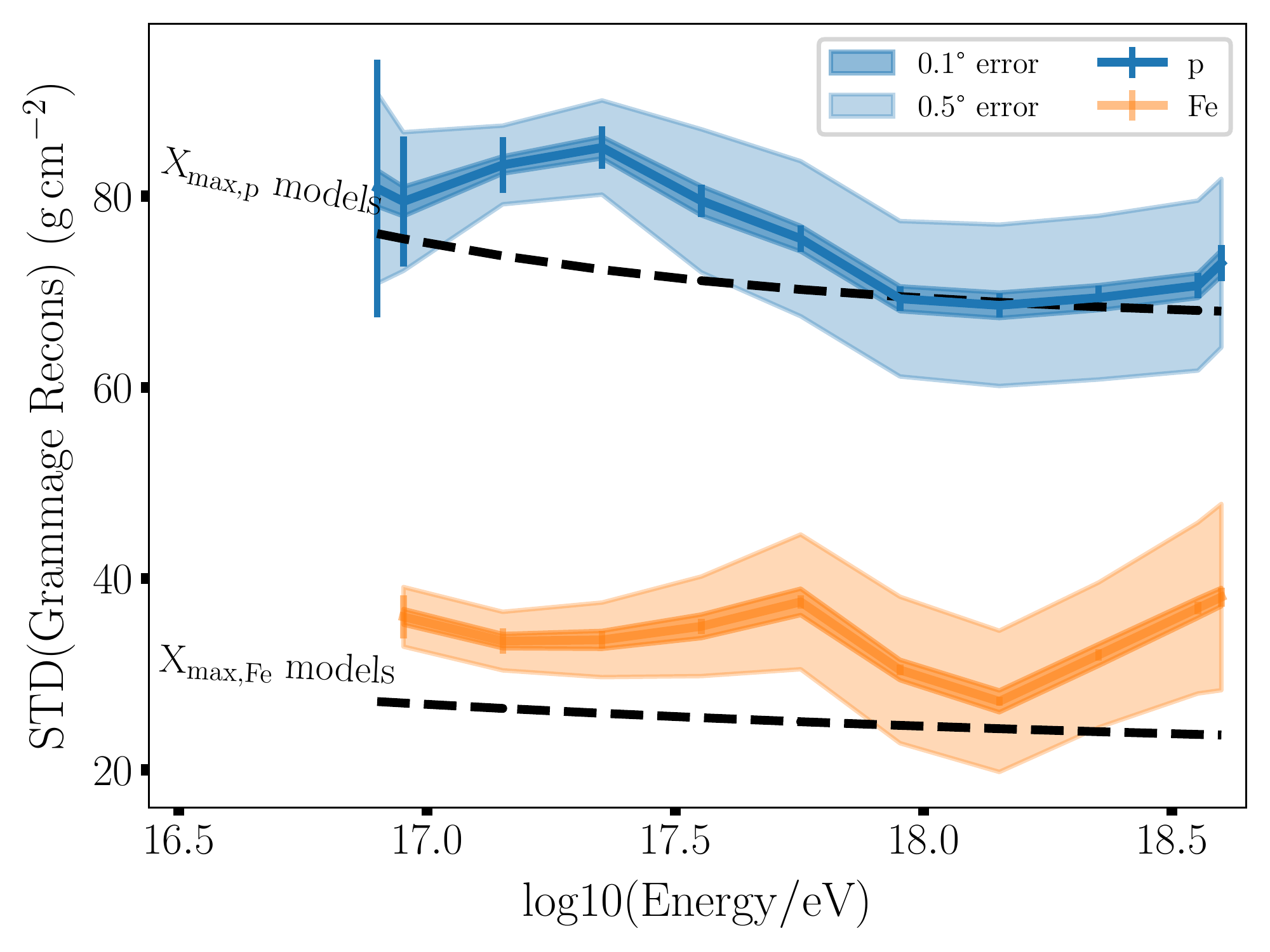}
    \caption{Expected standard deviation of the reconstructed grammage for proton (blue) and iron (orange) primaries. As for figure \ref{fig:radiogram_mean}, the error-bars show the statistical fluctuations, taken equal to $\sigma_{X_{\rm e}}/ \sqrt{N}$ where N is the number of simulated showers per energy slice. The width of the shaded areas correspond to the standard deviation of the $\sigma_{X_{\rm e}}^{i=1..100}$ distribution computed for errors on angular reconstruction of 0.1 (dark) and 0.5{\degree} (light) (see text for details). The black dotted lines correspond to the standard deviation of the $X_{\rm max}$ distributions from the proton and iron simulation data sets used in this study. At energies below $\sim10^{16.7}$\,eV the standard deviation increases drastically for iron primaries as the number of antennas then becomes very small. The effect is shifted to lower energies for protons because they are more penetrating and thus induce a stronger radio signal on ground for a same energy.}
    \label{fig:radiogram_std}
\end{figure}

We also observe in Figure \ref{fig:radiogram_mean} that radio grammage is $\sim$80 to $\sim$120\,g\,cm$^{-2}$ larger for proton primaries than iron ones over the energy range considered. This difference remains significant even when errors on the reconstruction of the shower direction of origin are taken into account, an effect estimated by applying the following treatment: 
\begin{enumerate}[i)]
    \item For each shower in the simulation set, a $realized$ direction of origin is drawn in a $\sigma$=0.1{\degree} gaussian distribution around the true direction to account for reconstruction errors. This resolution may be achievable by extended radio arrays~\cite{Valentin_PhD}, but a conservative value $\sigma$=0.5{\degree} is also considered for the sake of completeness. The corresponding radio grammage $X_{\rm e}^i$ is then computed. 
    \item Once the process is completed for all showers in the simulation dataset, the average value $\expval{X_{\rm e}}^i$ and standard deviation $\sigma_{X_{\rm e}}^i$ are computed from the full $X_{\rm e}^i$ distribution of realisation $i$. 
    \item Steps i) and ii) are repeated 100 times. The standard deviations of the $\expval{X_{\rm e}}^{i=1..100}$ and $\sigma_{X_{\rm e}}^{i=1..100}$ distributions for the 100 realistic realisation are taken as the respective contributions of the angular reconstruction error to the systematic uncertainties on the average and standard deviation of the radio grammage. The corresponding $\pm$1-$\sigma$ error bands are shown as shaded areas on Figures \ref{fig:radiogram_mean} and \ref{fig:radiogram_std} respectively.
\end{enumerate}
It can be observed from Figure \ref{fig:radiogram_mean} that the effect of a 0.1{\degree} resolution on the computation of the mean radio grammage is negligible. A 0.5{\degree} value yields a $\pm$10\,g\,cm$^{-2}$ uncertainty, and thus still allows for a precise computation of grammage from the radio source position. It also causes a $\sim$10-20\,g\,cm$^{-2}$ offset towards larger values of $\expval{X_{\rm e}}$, because an error $+ \delta \theta$ on directions towards the horizon induces a (positive) bias on radio grammage larger than the (negative) bias resulting from an error $- \delta \theta$ of the same absolute value towards the zenith. This effect increases with zenith angle, which explains why the bias is negligible at energies below 10$^{17.5}$\,eV, when only the less-inclined showers can be detected. 

The standard deviation of the radio grammage distribution ($\sigma_{X_{\rm e}}$) is plotted in Figure \ref{fig:radiogram_std} for the same energy slices for proton and iron primaries. As for its mean value, this quantity differs for iron and proton primaries, with values slightly larger --by $\sim$10\,g\,cm$^{-2}$ at most-- than the intrinsic fluctuations of $\vec{X}_{\rm max}$ ($\sigma_{X_{\rm max}}$). This could mean that the resolution of the method itself is negligible compared to shower-to-shower fluctuations, or that the fluctuations of $\vec{X}_{\rm e}$ are smaller than those of $X_{\rm max}$. One however has to be cautious with the latter explanation, again because the emission point $\vec{X}_{\rm e}$ is only a construction arising from the spherical model, and may not correspond to a physical reality. 

As there is no such thing as a "true" value for $\vec{X}_{\rm e}$,  we evaluate the resolution of our method for primary identification by computing its figure of merit (FOM) defined in \cite{Holt:2019fnj} as\,:
\begin{equation}
\label{eq:fom}
FOM = \frac{\lvert \mu_{Fe} - \mu_p \rvert}{\sqrt{\sigma_{Fe}^2+\sigma_p^2}}    
\end{equation}
where $\mu$ and $\sigma$ are the mean value and standard deviation for radio grammage respectively, computed here for proton and iron primaries. \\
The resulting value ranges between 1.0 and 1.4 between $10^{17}$ and $10^{18.5}$\,eV, without significant correlation with energy. This is somewhat lower than the $\sim$1.4 value computed over this energy range for $X_{\rm max}$ from the model lines shown in Figures \ref{fig:radiogram_mean} and \ref{fig:radiogram_std} for $\mu_{\rm X_{max}}$ and $\sigma_{\rm X_{max}}$, where only shower-to-shower fluctuations are accounted for. When a 17\,g/cm$^2$ value\,\cite{2014PhRvD..90h2003B} is added quadratically to $\sigma_{\rm X_{max}}$ to account for reconstruction resolution, the $X_{\rm max}$ FOM is slightly degraded to 1.3, a value still better than radio grammage on average. Hence our method does not seem fully competitive with $X_{\rm max}$ for shower composition studies and a clear identification of the nature of the primary will be hardly achievable on a shower-to-shower basis with this method, at least for the radio frequency range considered in the present study.
The effect of various systematic effects  above-mentioned -- realistic experimental conditions on trigger timing, frequency range of the signal filtering, trigger and detector layout, etc.--- yet remains to be studied before a precise conclusion can be made on the potential of this parameter for composition studies.

\section{Conclusion}

We have shown in this article that the radio wavefront of air showers can be adequately described by a simple sphere model (accounting for the anisotropy of the atmosphere) for zenith angles larger than 60$\degree$.  Thanks to the very large extension of the radio footprint at ground, the position of the center of the sphere realtive to the shower axis can be reconstructed with remarkable precision on very large arrays, even if the association of the position of the center of the sphere with a physical emission point is uncertain at this stage. This implies that if an additional point on the shower axis were to be determined (e.g. the position of the shower core on ground), the shower incoming direction could be reconstructed with good accuracy. This idea was briefly presented in \cite{Decoene:2021ncf} and will also be the topic of a future article.

Our treatment also shows that the depth in atmosphere of the source position significantly differs for showers induced by protons and irons, even when systematic errors on the reconstruction of the shower direction of origin are taken into account. This hints towards the possibility to use the radio source position as proxy for the study of cosmic ray composition.

Additional studies on possible sources of systematic effects need to be done to assess this.
The present study of principle will thus be completed by taking into account realistic experimental conditions and systematic effects. 
\subsection*{Acknowledgments}
We thank our colleagues from the GRAND collaboration ---and in particular Krijn de Vries, Simon Chiche and Simon Prunet--- for their valuable comments and suggestions throughout this study. We also thank the reviewers for their valuable comments which helped improving the quality of this article.

This work was supported by the Programme National des Hautes Energies of CNRS/INSU with INP and IN2P3, co-funded by CEA and CNES. The simulations presented in this paper were produced at the IN2P3/CNRS computing center.

\bibliographystyle{elsarticle-num} 
\bibliography{references.bib} 

\end{document}